\documentclass{article}
\usepackage{authblk}
\usepackage[utf8]{inputenc}
\usepackage{amsmath,amssymb,amsfonts,amsthm}
\usepackage{physics}
\usepackage{dsfont}
\usepackage{graphicx}
\usepackage{xcolor}
\usepackage{cancel}
\usepackage[numbers,sort&compress,square]{natbib}
\usepackage{color}
\usepackage{bbm}
\usepackage[linktocpage]{hyperref}
\usepackage{appendix}
\hypersetup{colorlinks=true,citecolor=blue,linkcolor=blue, urlcolor=blue, breaklinks=true}
\usepackage[eulergreek]{sansmath}
\usepackage{mathtools}
\usepackage{scalerel}

\usepackage{ulem}

\usepackage{bm} 
\usepackage{subfigure} 
\usepackage{environ}
\usepackage{url}
\usepackage{hyperref}
\usepackage[margin=1in]{geometry}

\usepackage{environ}
\NewEnviron{eqs}{%
\begin{equation}\begin{split}
    \BODY
\end{split}\end{equation}}

\newcommand{\ZZ}{{\mathbb Z}}
\newcommand{\Z}{{\mathbb Z}}

\newcommand{\eps}{\epsilon}

\graphicspath{{./Figures/}}

\begin{document}

\title{Higher-Form Anomalies on Lattices}

\author[1, *]{Yitao Feng}
\author[2, *, $\dagger$]{Ryohei Kobayashi}
\author[1, $\ddagger$]{Yu-An Chen}
\author[3, $\#$]{Shinsei Ryu}

\affil[1]{International Center for Quantum Materials, School of Physics, Peking University, Beijing 100871, China}
\affil[2]{School of Natural Sciences, Institute for Advanced Study, Princeton, NJ 08540, USA}
\affil[3]{Department of Physics, Princeton University, Princeton, NJ 08544, USA}

\renewcommand*{\thefootnote}{*}
\footnotetext[1]{These authors contributed equally to this work.}
\renewcommand*{\thefootnote}{$\dagger$}
\footnotetext[1]{Contact author: ryok@ias.edu}
\renewcommand*{\thefootnote}{$\ddagger$}
\footnotetext[1]{Contact author: yuanchen@pku.edu.cn}
\renewcommand*{\thefootnote}{$\#$}
\footnotetext[1]{Contact author: shinseir@princeton.edu}

\renewcommand{\thefootnote}{\arabic{footnote}}

\date{\today}

\maketitle

\begin{abstract}
Higher-form symmetry in a tensor product Hilbert space is always emergent: the symmetry generators become genuinely topological only when the Gauss law is energetically enforced at low energies. 
In this paper, we present a general method for defining the ’t Hooft anomaly of higher-form symmetries in lattice models built on a tensor product Hilbert space.
In (2+1)D, for given Gauss law operators realized by finite-depth circuits that generate a finite 1-form $G$ symmetry, we construct an index representing a cohomology class in $H^4(B^2G, U(1))$, which characterizes the corresponding ’t Hooft anomaly. This construction generalizes the Else–Nayak characterization of 0-form symmetry anomalies.
More broadly, under the assumption of a specified formulation of the $p$-form $G$ symmetry action and Hilbert space structure in arbitrary $d$ spatial dimensions, we show how to characterize the ’t Hooft anomaly of the symmetry action by an index valued in $H^{d+2}(B^{p+1}G, U(1))$.
\end{abstract}

\tableofcontents

\section{Introduction}

Symmetry plays a central role in the study of quantum many-body systems and quantum field theory. Beyond conventional 0-form global symmetries, which act on point-like excitations, recent developments have revealed the importance of higher-form symmetries that act on extended operators such as lines, surfaces, or higher-dimensional objects~\cite{Gaiotto:2014kfa}. Higher-form symmetries are ubiquitous in gauge theories, various models of spin liquids and quantum codes, and constrain their low-energy dynamics.

An important subtlety arises when realizing higher-form symmetries in lattice models with tensor product Hilbert spaces. In such settings, higher-form symmetry generators are not strictly topological operators at the microscopic level. Instead, they are emergent and become topological operators only in the presence of an energetically enforced Gauss law at low energies. This emergent nature complicates the definition and characterization of ’t Hooft anomalies. An ’t Hooft anomaly refers to an obstruction to gauging a global symmetry, and its presence implies that the system cannot be realized in a short-range entangled (SRE) phase \cite{chen2010}. For instance, the Lieb-Schultz-Mattis theorem~\cite{LSM1961,oshikawa2,hastings2004} and its generalizations~\cite{cheng2016set, cho2017anomaly, Kobayashi2019lsm, else2020lsm, prem2020lsm, seiberg2022lsm, Kapustin2025LSM, Kawabata2024LSM, Gu2024LSMopen, Liu2024spinS, Pace2025LSM} encode non-trivial constraints on the low-energy physics of lattice systems that originate from mixed anomalies between spatial and internal symmetries.
Anomalies also put strong constraints on deconfinement in gauge theories~\cite{shimizu2018,hsin2020se,seifnashri2021sym} and imply the nontrivial edge states of symmetry protected topological phases~\cite{Chen2011Twodimensional, chen2013SPT,ElseNayak2014,tiwari2018}.
't Hooft anomalies of higher-form symmetries have recently been used to constrain the entanglement structure of both pure and mixed states, providing refined dynamical constraints on the system \cite{Lessa:2024wcw, li2024anyon, lessa2025higher, zhou2025finiteT, Hsin:2025pqb}.

Previous work has clarified the structure of anomalies for ordinary (0-form) global symmetries. In continuum quantum field theory (QFT), ’t Hooft anomalies of bosonic systems are classified by group cohomology. On the lattice, Else and Nayak developed a characterization of 0-form symmetry anomalies by group cohomology directly within a tensor product Hilbert space on the lattice \cite{ElseNayak2014}, establishing a concrete connection to the continuum classification of 't Hooft anomalies.
In particular, for a given symmetry operator generated by a finite-depth circuit in a (1+1)D lattice model, their construction defines an anomaly index valued in $H^3(BG,U(1))$, which matches the continuum classification of bosonic 0-form anomalies.
Their construction, originally formulated for generic finite-depth circuits in (1+1)D, has recently been generalized to 0-form symmetries in (2+1)D \cite{kawagoe2025anomaly, kapustin2025anomaly2d}. 

For higher-form symmetries, recent progress has been made in Ref.~\cite{kobayashi2025generalizedstatistics}, which characterizes ’t Hooft anomalies using Berry phase invariants defined by the action of symmetry operators acting on a reference symmetric state. We also note that Ref.~\cite{kapustin2025higher} discusses anomalies of higher symmetries in lattice gauge theory\footnote{Ref.~\cite{kapustin2025higher} assumes that the 1-form symmetry operators can be truncated while still commuting with Gauss laws. In our setting, described in the next section, this condition is not required.} using crossed squares through homotopy theory and operator algebras.
The Berry phase invariants developed in Ref.~\cite{kobayashi2025generalizedstatistics} generalize the ``T-junction'' invariant defined in Ref.~\cite{Levin2003Fermions} for 1-form symmetry in (2+1)D.
Meanwhile, it is widely expected that anomalies should admit a formulation entirely in terms of symmetry operators themselves, without reference to any state in the Hilbert space. From this perspective, a comprehensive framework for higher-form symmetry anomalies in lattice models remains less established.

In this work, we develop such a framework; we introduce a general method to define the ’t Hooft anomaly of higher-form symmetries in tensor product Hilbert spaces. In (2+1)D, given a set of Gauss law operators implemented by finite-depth circuits that generate a 1-form $G$ symmetry, we define an index valued in the cohomology class $H^4(B^2G, U(1))$, which characterizes the anomaly. This construction naturally generalizes the Else–Nayak approach. 
We show that the anomaly index indeed defines an obstruction to symmetric SRE states, therefore gives a microscopic definition of 1-form 't Hooft anomaly. 
Furthermore, by assuming a specific structure for the action of a $p$-form $G$ symmetry and the underlying Hilbert space, we extend the characterization to arbitrary spacetime dimension $(d+1)$, where the anomaly is captured by an index in $H^{d+2}(B^{p+1}G, U(1))$.
This provides a unified framework for diagnosing higher-form symmetry anomalies in lattice models, establishing a direct correspondence between higher-form anomalies in microscopic lattice models and cohomological classifications of QFT anomalies.

This paper is organized as follows. In Sec.~\ref{sec:1form}, we introduce the 1-form symmetry in (2+1)D lattice models generated by finite-depth circuits, and define an $H^4$ index that characterizes the 't Hooft anomaly of the given 1-form symmetry. In Sec.~\ref{sec:higher}, we generalize the method to the higher-form symmetries in generic spacetime dimensions, assuming a specific form of the tensor product Hilbert space and symmetry actions. In Sec.~\ref{sec:statistics}, we comment on the relations between the anomaly index and the T-junction invariants in (2+1)D discussed in Ref.~\cite{Levin2003Fermions}.

\section{1-form symmetry in (2+1)D}
\label{sec:1form}

\subsection{1-form symmetry on lattices}

Consider a lattice quantum system on a tensor product Hilbert space in two spatial dimensions. In this section, we define finite 1-form symmetry in the most generic setup, where the symmetry is generated by a finite-depth circuit supported at a codimension-1 locus of the space.

\begin{figure}[tbh]
\centering
\includegraphics[width=0.8\textwidth]{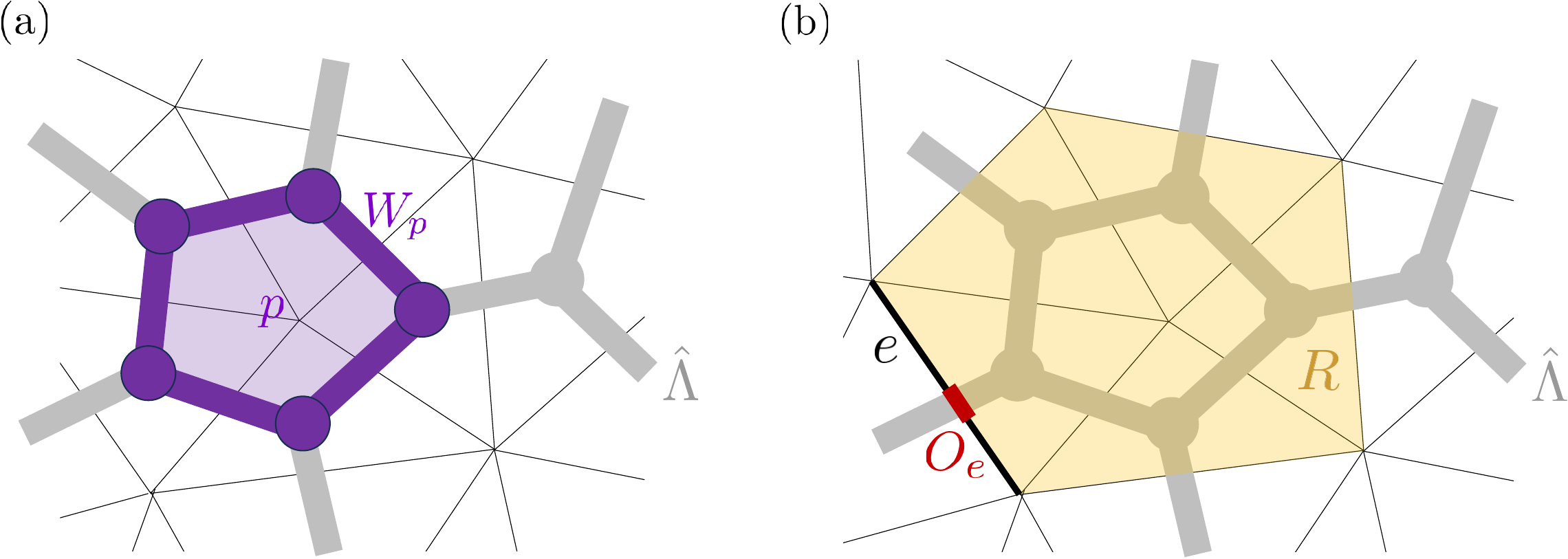}
\caption{(a): The symmetry operators are supported at the thickened dual lattice of a mesoscopic triangulation of a 2d lattice system. $W_p$ is supported at a closed loop along the boundary of a plaquette $p$. A plaquette of the dual lattice $p$ corresponds to a vertex in the original lattice $\Lambda$. 
(b): The local operator $O_e$ is supported at the intersection between an edge $e$ of $\partial R$ and an edge of $\hat{\Lambda}$.
}
\label{fig:duallattice}
\end{figure}

We consider a ``mesoscopic'' triangulation $\Lambda$ of the 2d space; the size of each edge in the triangulation is taken much larger than the locality length and the circuit depth of symmetry operators.
All symmetry operators will be finite-depth circuits, supported within the thin strip along dual lattice $\hat{\Lambda}$ of the triangulation. See Fig.~\ref{fig:duallattice}(a).

The 1-form symmetry is defined through the Gauss laws. The Gauss law operators $W_p$ are defined at each plaquette $p$ of the mesoscopic lattice $\hat{\Lambda}$. This is a finite-depth unitary supported at the plaquette boundary $\partial p$. The Gauss law condition is then $W_p=1$ for any plaquettes, which makes the symmetry operator topological. We require the following conditions on $W_p$:
\begin{enumerate}
    \item For 1-form $G$  symmetry, there is a Gauss law operator $W_p^{(g)}$ labeled by a group element $g\in G$ for all plaquettes $p$, satisfying the group algebra on each plaquette: 
    \begin{align}
        W^{(g)}_p W^{(g')}_p = W^{(gg')}_p~,
        \label{eq:W follows group}
    \end{align}
    for $g,g'\in G$. Concretely, for a generic finite Abelian group $G=\bigoplus_j \Z_{N_j}$, we have a set of Gauss law operators $\{W^{(j)}_p\}$ on each plaquette $p$ labeled by $\{j\}$, where $[W^{(j)}_p,W^{(j')}_p]=1$ with a group commutator $[U,V]:=U^{-1}V^{-1}UV$, and $(W^{(j)}_p)^{N_j}=1$.

    \item $W_p$ on different plaquettes are commutative: $[W_p^{(g)}, W_{p'}^{(g')}]=1$ for generic $p,p',g,g'$. This ensures that the Gauss law constraint $\{W^{(g)}_p=1\}_{g,p}$ has a solution.
    \item The product of $W^{(j)}_p$ over the whole space becomes identity:
    \begin{align}
        \prod_{p} W^{(j)}_p =1~.
        \label{eq: product of Gauss laws}
    \end{align}
    This ensures that the operators $W^{(j)}_p$ deform a line operator generating symmetry into another line operator, therefore the symmetry operators become topological when the Gauss law constraint $\{W^{(g)}_p=1\}_{g,p}$ is enforced.
    
\end{enumerate}

This completes the definition of a 1-form $G$  symmetry on a 2d lattice system with a finite Abelian group $G$. For instance, the symmetry generator for $j$-th generator of $G=\oplus_j \Z_{N_j}$ at a closed loop is given by
\begin{align}
    W^{(j)}(\partial \hat{R}) =  \prod_{p\in \hat{R}} W^{(j)}_p~,
\end{align}
where $\hat{R}$ is a disk region formed by a collection of plaquettes $p$.
Note that due to the property \eqref{eq: product of Gauss laws}, the product of Gauss laws inside the disk $\hat{R}$ cancels out. 

\subsection{Else-Nayak type index}
\label{subsec:2+1dindex}
\begin{figure}[tbh]
\centering
\includegraphics[width=0.3\textwidth]{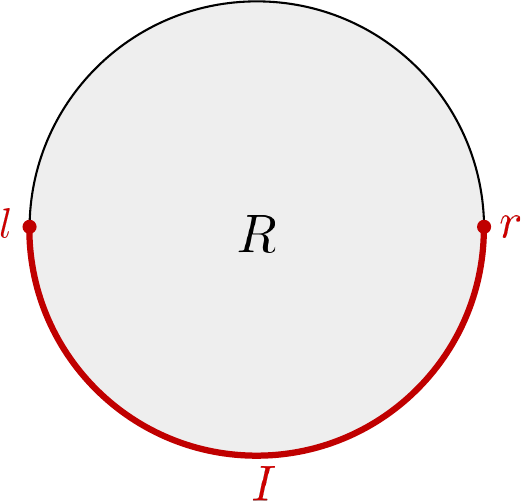}
\caption{An interval $I$ at the boundary of the region $R$.}
\label{fig:region}
\end{figure}

Let $G$ be a finite Abelian group, $G=\oplus_j \Z_{N_j}$.
Here, we define a 4-cocycle $\omega\in H^4(B^2G, U(1))$ out of the lattice 1-form $G$ symmetry defined above. 
This is done in a similar manner to the case of 0-form symmetry anomaly in (2+1)D~\cite{kawagoe2025anomaly, kapustin2025anomaly2d}, as a generalization of the anomaly index defined by Else and Nayak \cite{ElseNayak2014}. See Appendix \ref{app:elsenayak} for a review of the Else-Nayak index of 0-form symmetry in (1+1)D.
We first define a symmetry operator $U(\epsilon)$ labeled by a 0-cochain $\epsilon\in C^0(\Lambda,G)$ of the triangulation,
\begin{align}
    U(\epsilon) = \prod_{p} (W_p)^{\epsilon(p)}~,
    \quad \text{with} \quad 
    (W_p)^{\epsilon(p)} := \prod_j (W_p^{(j)})^{\epsilon_j(p)}~,
\end{align}
where we decompose $\eps = \oplus_j \eps_j$, with each $\eps_j \in C^0(\Lambda,\ZZ_{N_j})$. 
Since $\prod_p W^{(g)}_p=1$ for any group labels $g\in G$, $U(\epsilon)$ satisfies
\begin{align}
    U(\epsilon + g) = U(\epsilon)~,
\end{align}
with a constant 0-form $g\in C^0(\Lambda,G)$ labeled by $g\in G$. 
Below, we illustrate the procedure to define $\omega\in H^4(B^2G,U(1))$ by steps:

\begin{enumerate}
    \item 
Let us consider a disk region $R$ of the triangulation $\Lambda$. We take the size of $R$ to be large in the mesoscopic lattice $\Lambda$. 
$R$ is a collection of 2-simplices inside a disk region, see Fig.~\ref{fig:duallattice} (b).
Since $U(\epsilon)$ is a finite-depth circuit, one can choose a restriction of $U(\epsilon)$ within $R$, which we denote by $U_R(\epsilon)$. In particular, we consider the following form of $U_R(\epsilon)$,
\begin{align}
    U_R(\epsilon) = \prod_{p\in \partial R} (W_{p;R})^{\epsilon(p)}\prod_{p\in\text{Int}(R)} (W_p)^{\eps(p)}~,
    \label{eq:explicit UR}
\end{align}
where $p\in R$ denotes the plaquettes along the boundary $\partial R$, and we choose a restriction $W_{p;R}$ of the Gauss law operator. $p\in\text{Int}(R)$ denotes the plaquettes inside the region, where $W_p$'s are not truncated.
A pair of operators $W_{p;R}$ at neighboring plaquettes along the boundary $\partial R$ no longer commute with each other. Therefore, we are fixing an ordering of operators to define the product $\prod_{p\in \partial R}$.\footnote{Due to mesoscopic nature of the lattice $\hat\Lambda$, $U_R(\epsilon)$ is a finite-depth circuit with any choice of ordering in the product.} Later in Sec.~\ref{subsec:redundancy}, we will see that the anomaly index is independent of possible ambiguities to define $U_R(\eps)$, i.e., the ordering in the product and choices of truncations to define $W_{p;R}$.

We then define an operator 
\begin{align}
    \Omega(\epsilon_{01},\epsilon_{12}, g_{012}) =U_R(\epsilon_{01})U_R(\epsilon_{12})    U_R(\epsilon_{01}+\epsilon_{12}-g_{012})^{-1}~,
    \label{eq:defineOmega}
\end{align}
which has a support within a thin strip along $\partial R$. Here we note that the subscripts $ \epsilon_{jk}, g_{jkl} $ are not related to the simplices of $ \Lambda $ or to any 2d space. Rather, they serve as fictitious labels. As we will see in Sec.~\ref{sec:graphical} and Sec.~\ref{sec:higher}, these subscripts are useful for providing a graphical representation of the procedure and for facilitating generalizations to higher dimensions.

This operator $\Omega$ is expressed as a product of separate local operators:
\begin{align}
     \Omega(\epsilon_{01},\epsilon_{12}, g_{012}) = \prod_{e\in\partial R} O_{e}(\epsilon_{01}, \epsilon_{12}, g_{012})~,
     \label{eqref:Oe}
\end{align}
where $e$ is an edge of the original mesoscopic lattice $\Lambda$ along $\partial R$. Each $O_e$ is supported at an intersection between an edge $e\in \partial R$ of the triangulation $\Lambda$ and an edge of $\hat{\Lambda}$ (the intersection is shown as a red region in Fig.~\ref{fig:duallattice} (b)).
Due to the form of $U_R(\epsilon)$ in \eqref{eq:explicit UR} which is local with respect to $\{\epsilon\}$, each operator $O_{e}(\epsilon_{01},\epsilon_{12}, g_{012})$ is a local functional of 0-forms $\epsilon_{01},\epsilon_{12}$. We note that ``locality'' here is purely defined on a mesoscopic triangulation $\Lambda$, instead of the original locality scale of the lattice model or circuits; it is a local functional in the sense that $O_e(\{\epsilon\})$ only depends on the values $\epsilon(v)$ at vertices $v\in\partial e$ on $\Lambda$.

\item 
Let us introduce a shorthand notation $\epsilon_{ik} := \epsilon_{ij}+\epsilon_{jk}-g_{ijk}$, with $i<j<k$. The operator $\Omega$ satisfies
\begin{align}
    \Omega(\epsilon_{01},\epsilon_{12})\Omega(\epsilon_{02},\epsilon_{23})
    =~^{\epsilon_{01}}\Omega(\eps_{12},\eps_{23}) \Omega(\eps_{01},\eps_{13})~,
    \label{eq:Omega cocycle}
\end{align}
where $^{\epsilon}O = U_R(\epsilon) O U_R(\epsilon)^{-1}$, and we suppressed the dependence of $\Omega$ on $\{g\}$ in the expression. 
Due to the above form of $\Omega$, this implies that the following combination becomes an overall phase
\begin{align}
    e^{2\pi i F_e(\epsilon_{01},\eps_{12},\eps_{23},\{g\})}:= O_e(\epsilon_{01},\epsilon_{12})O_e(\epsilon_{02},\epsilon_{23})\left(^{\epsilon_{01}}O_e(\eps_{12},\eps_{23})O_e(\eps_{01},\eps_{13})\right)^{-1}~.
    \label{eq:O cocycle}
    \end{align}
$F_e(\epsilon_{01},\eps_{12},\eps_{23},\{g\})\in\mathbb{R}/\mathbb{Z}$ is a local functional of 0-forms $\epsilon_{01},\epsilon_{12},\epsilon_{23}$ on the mesoscopic triangulation $\Lambda$, and also depends on group labels $g_{ijk}$ with $i,j,k\in\{0,1,2,3\}$. These group labels satisfy
\begin{align}
    g_{013} + g_{123} -g_{012} - g_{023} = 0~.
\end{align}
Since $F_e$ is defined on each edge of $\partial R$, this is interpreted as a 1-cocycle $F\in Z^1(\partial R,\mathbb{R}/\mathbb{Z})$. Due to \eqref{eq:Omega cocycle}, its integral over $\partial R$ becomes trivial,
\begin{align}
    \int_{\partial R}F_e(\epsilon_{01},\eps_{12},\eps_{23},\{g\})=\sum_{e\in\partial R}F_e(\epsilon_{01},\eps_{12},\eps_{23},\{g\}) = 0 \quad \text{mod $1$.}
    \end{align}
This implies that $F$ is a coboundary on $\partial R$, $F=dA$ with some 0-cochain $A(\epsilon_{01},\eps_{12},\eps_{23},\{g\})\in C^{0}(\partial R, \mathbb{R}/\mathbb{Z})$, supported on vertices of $\Lambda$ along $\partial R$. 

Now, let us introduce a restriction of $\Omega(\epsilon_{01},\epsilon_{12})$ within an interval $I$ of $\partial R$ ($I$ is a collection of edges $e$ of $\Lambda$),
\begin{align}
     \Omega_I(\epsilon_{01},\epsilon_{12}) = \prod_{e\in I} O_{e}(\epsilon_{01}, \epsilon_{12})~.
 \end{align}
 Then we can extract the form of $A$ using $\Omega_I$,
\begin{align}
\begin{split}
 \Omega_I(\epsilon_{01},\epsilon_{12})\Omega_I(\epsilon_{02},\epsilon_{23}) (^{\epsilon_{01}}\Omega_I(\epsilon_{12},\epsilon_{23})\Omega_I(\epsilon_{01},\epsilon_{13}))^{-1} = e^{2\pi i \int_{I}F_e(\epsilon_{01},\eps_{12},\eps_{23},\{g\})} = e^{2\pi i (A_l-A_r)}~,
\end{split}
\label{eq:Gamma}
\end{align}
where $l,r$ are two vertices at the ends of the interval $I$, see Fig.~\ref{fig:region}.

 Now we define an anomaly index $\omega_l$ using a functional $A_l$. First, the functional $F_e$ satisfies the ``3-cocycle condition'',
 \begin{align}
     F_e(\epsilon_{01},\epsilon_{12},\epsilon_{23}) + F_e(\epsilon_{01},\epsilon_{13},\epsilon_{34}) + F_e(\epsilon_{12},\epsilon_{23},\epsilon_{34}) = F_e(\epsilon_{02},\epsilon_{23},\epsilon_{34}) + F_e(\epsilon_{01},\epsilon_{12},\epsilon_{24}) \quad \text{mod $1$,}
     \label{eq:3cocycle Fe}
\end{align}
where we suppressed the dependence on group labels $\{g\}$.
This is derived by rewriting the operator $O_e(\eps_{01},\eps_{12})O_e(\eps_{02},\eps_{23})O_e(\eps_{03},\eps_{34})$ using \eqref{eq:O cocycle} in two different ways. One one hand we get
\begin{align}
\begin{split}
    O_e(\eps_{01},\eps_{12})O_e(\eps_{02},\eps_{23})O_e(\eps_{03},\eps_{34}) =& e^{2\pi i( F_e(\epsilon_{01},\epsilon_{12},\epsilon_{23}) + F_e(\epsilon_{01},\epsilon_{13},\epsilon_{34}) + F_e(\epsilon_{12},\epsilon_{23},\epsilon_{34}))} \\  
&\times (^{\eps_{01}}(^{\eps_{12}}O_e(\eps_{23},\eps_{34})))(^{\eps_{01}}O_e(\eps_{12},\eps_{24}))O_e(\eps_{01},\eps_{14})~.
\end{split}
\end{align}
On the other hand we get
\begin{align}
\begin{split}
    O_e(\eps_{01},\eps_{12})O_e(\eps_{02},\eps_{23})O_e(\eps_{03},\eps_{34}) =& e^{2\pi i(F_e(\epsilon_{02},\epsilon_{23},\epsilon_{34}) + F_e(\epsilon_{01},\epsilon_{12},\epsilon_{24}) )} \\  
&\times O_e(\eps_{01},\eps_{12}) (^{\eps_{02}}O_e(\eps_{23},\eps_{34}))O_e(\eps_{01},\eps_{12})^{-1} \\
&\times (^{\eps_{01}}O_e(\eps_{12},\eps_{24}))O_e(\eps_{01},\eps_{14})~.
\end{split}
\end{align}
Due to the definition of $O_e$, we have $(^{\eps_{01}}(^{\eps_{12}}O_e(\eps_{23},\eps_{34})))=O_e(\eps_{01},\eps_{12}) (^{\eps_{02}}O_e(\eps_{23},\eps_{34}))O_e(\eps_{01},\eps_{12})^{-1}$. Therefore equating the above two equations leads to \eqref{eq:3cocycle Fe}.

Since $F=dA$ satisfies the 3-cocycle condition \eqref{eq:3cocycle Fe}, $A_l$ also follows a similar relation. This leads to the following definition of the 4-cocycle $\omega_l$ as
\begin{align}
    \begin{split}
\omega_l(\epsilon_{01},\epsilon_{12},\epsilon_{23},\epsilon_{34},\{g\}) &:= A_l(\epsilon_{01},\epsilon_{12},\epsilon_{23}) + A_l(\epsilon_{01},\epsilon_{13},\epsilon_{34}) + A_l(\epsilon_{12},\epsilon_{23},\epsilon_{34})  \\
&\quad 
- A_l(\epsilon_{02},\epsilon_{23},\epsilon_{34}) - A_l(\epsilon_{01},\epsilon_{12},\epsilon_{24})~,
\label{eq:omega def}
\end{split}
\end{align}
where $\omega_l$ depends on $g_{jkl}$ with $j,k,l\in\{0,1,2,3,4\}$.
Using the right endpoint $r$ of $I$, one can define the other index $\omega_r(\epsilon_{01},\epsilon_{12},\epsilon_{23},\epsilon_{34},\{g\})$ by the same method.

\end{enumerate}

\subsubsection{Index defines $[\omega]\in H^4(B^2G,U(1))$}

Let us describe a number of important properties satisfied by the above index $\omega_l$, showing that the anomaly index defines an element of $H^4(B^2G,U(1))$. 
\begin{enumerate}
    \item First, $\omega_l(\epsilon_{01},\epsilon_{12},\epsilon_{23},\epsilon_{34},\{g\})$ is independent of 0-forms $\{\epsilon\}$, meaning that it is a function of $\{g\}$ alone.
We can see this from a ``3-cocycle condition'' satisfied by $F_e$ in \eqref{eq:3cocycle Fe}. 
This implies that 
\begin{align}
    \omega_l(\epsilon_{01},\epsilon_{12},\epsilon_{23},\epsilon_{34},\{g\}) -\omega_r(\epsilon_{01},\epsilon_{12},\epsilon_{23},\epsilon_{34},\{g\}) = 0 \quad \text{mod 1}.
    \label{eq:independence of omega}
\end{align}
Since $\omega_r$ is a local function depending only on ${\epsilon}$ at the right end $r$,  $\omega_l = \omega_r$ is independent of ${\epsilon}$ at the left end $l$. This shows that $\omega_l$ is independent of $\{\epsilon\}$. From now, we simply write $\omega_{01234}:=\omega_l(\epsilon_{01},\epsilon_{12},\epsilon_{23},\epsilon_{34},\{g\}) =\omega_l(g_{012},\dots, g_{234})$, which is a function of $g_{jkl}$ with $j,k,l\in\{0,1,2,3,4\}$. This defines an element $\omega\in C^4(B^2G,U(1))$. See Appendix \ref{app:EMspace} for a review of Eilenberg-MacLane spaces $B^{p}G$ and their cohomology.

\item 
$\omega$ satisfies the cocycle condition, meaning that  
\begin{align}
    \omega_{12345} + \omega_{01345} + \omega_{01235} - \omega_{02345} - \omega_{01245} - \omega_{01234} = 0 \quad \text{mod 1},
\end{align}
where $\omega_{ijklm} = \omega(g_{ijk},\dots, g_{klm})$. This 4-cocycle condition is shown by rewriting the following combination of $A_l$'s in two different ways,
\begin{align}
    A_l(\eps_{01},\eps_{12},\eps_{23}) + A_l(\eps_{01},\eps_{13},\eps_{34}) + A_l(\eps_{12},\eps_{23},\eps_{34}) +  A_l(\eps_{01},\eps_{14},\eps_{45}) + A_l(\eps_{12},\eps_{24},\eps_{45}) + A_l(\eps_{23},\eps_{34},\eps_{45})~,
\end{align}
by a repeated use of \eqref{eq:omega def}.
Therefore $\omega$ defines an element $\omega\in Z^4(B^2G,U(1))$.

\item As we will shortly see in Sec.~\ref{subsec:redundancy}, by redefinitions of $U_R, O_e,A_l$, the 4-cocycle $\omega$ is shifted by $\omega+\delta \phi$ with $\phi\in C^3(B^2G,U(1))$. 
Here $\delta$ is a coboundary operation defined as
\begin{align}
    \delta\phi(g_{012},\dots, g_{234}) := \phi_{0123} - \phi_{0124} + \phi_{0134} - \phi_{0234} + \phi_{1234}~,
\end{align}
where $\phi\in C^3(B^2G,U(1))$ is a function of $\{g\}$, depending on $\{g\}$ through $\phi_{ijkl} = \phi(g_{ijk}, g_{jkl}, g_{ijl}, g_{ikl})$.
This leaves the cohomology class $[\omega]\in H^4(B^2G, U(1))$ invariant.

\end{enumerate}

\subsubsection{Invariance of the index}
\label{subsec:redundancy}
\paragraph{Phase redefinitions.}
During the above procedure to define the index $\omega$, there are a number of ambiguities to define the operators $O_e,A_l$ by phases. Here we describe such ambiguities and see how they affect the index:
\begin{enumerate}
    \item First, each operator $O_e(\epsilon_{01}, \epsilon_{12}, g_{012})$ in \eqref{eqref:Oe} at an edge $e=\langle vv'\rangle$ with a pair of vertices $v,v'$ of $\Lambda$ can be redefined by a phase $e^{2\pi id\eta_{e}} =e^{2\pi i(\eta_{v'}-\eta_v)}$. Here, $\eta$ is a 0-form $\eta\in C^0(\Lambda, U(1))$ which is a local functional of $\epsilon,g$. Such redefinitions do not shift $\Omega$, and preserve 
    the locality of $O_e$ with respect to $\{\epsilon\}$. 
    \item Due to the above redefinition, $A_l$ is shifted by a phase $\delta\eta_l$ with $\delta$ a coboundary operation defined as
\begin{align}
\begin{split}
    \delta\eta_l(\epsilon_{01},\epsilon_{12},\epsilon_{23},g_{012},g_{123},g_{013},g_{023})  := &-\eta_l(\epsilon_{01},\epsilon_{12}, g_{012}) -\eta_l(\epsilon_{01}+\epsilon_{12}- g_{012},\epsilon_{23}, g_{023}) \\
    &+ \eta_l(\epsilon_{12},\epsilon_{23}, g_{123}) + \eta_l(\epsilon_{01},\epsilon_{12}+\epsilon_{23}-g_{123}, g_{013})~.
    \end{split}
    \label{eq:chi}
\end{align}
Aside from the above phase redefinition, $e^{2\pi i A_l}$ can be redefined by a phase $e^{2\pi i\phi_{0123}}:= e^{2\pi i\phi(g_{012},g_{123},g_{013},g_{023})}$, 
which depends on $\{g\}$ but independent of $\{\epsilon\}$. 
Summarizing, the phase ambiguity of $A_l$ is
\begin{align}
    A_l\to A_l+ \delta\eta_l(\{\epsilon\},\{g\}) +\phi(\{g\})~.
    \label{eq: shift of A}
\end{align}

\item Due to the above redefinition, $\omega$ is shifted as $\omega \to \omega+\delta\phi$, with $\delta$ a coboundary operation defined as
\begin{align}
    \delta\phi(g_{012},\dots, g_{234}) := \phi_{0123} - \phi_{0124} + \phi_{0134} - \phi_{0234} + \phi_{1234}~.
    \label{eq:delta phi}
\end{align}
Note that $\omega$ is invariant under redefinitions by $\eta$.
This leaves the cohomology class $[\omega]\in H^4(B^2G,U(1))$ invariant. 

\end{enumerate}

\paragraph{Local operators.}
In addition to the above phase redefinitions, one can redefine the restricted symmetry operator $U_R(\epsilon)$ by a product of local operators at $\partial R$,
\begin{align}
    \tilde U_R(\epsilon) = \Sigma_\epsilon U_R(\epsilon)~,
    \label{eq:redefineSigma}
\end{align}
with $\Sigma_\epsilon$ a product of separate local operators at edges $e$ on $\partial R$,
\begin{align}
    \Sigma_\eps = \prod_{e\in\partial R} \Sigma_e(\eps)~,
\end{align}
where $\Sigma_e$ is a local functional of $\epsilon$, and has the same support as $O_e$, see Fig.~\ref{fig:duallattice} (b). Such redefinitions happen by choosing different restrictions of the Gauss law operators $W_{p;R}$, or changing the ordering of restricted Gauss law operators in the expression of \eqref{eq:explicit UR}.

To see this, it suffices to consider a simple case where $\Sigma_\eps$ is supported on a single edge $e$, 
$\Sigma_\eps= \Sigma_e(\eps)$. The index $\omega_l$ can be computed on an arbitrary vertex $l$ of $\partial R$, and due to \eqref{eq:independence of omega}, $\omega_l$ is independent of the choice of $l$. Let us choose $l$ to be away from the perturbation at $e$. Due to locality, such a perturbation cannot shift $A_l$ by a nontrivial functional of $\eps$ nearby the vertex $l$; its effect is at most shifting $A_l$ by $A_l\to A_l+\phi(\{g\})$ in \eqref{eq: shift of A}. Therefore, this again shifts $\omega$ by $\omega\to \omega+\delta \phi$, and leaves the cohomology class $[\omega]$ invariant.

\subsubsection{Remark on ``onsite'' anomalous symmetries} 
\label{subsec:onsite}
It is known that a 1-form symmetry in a lattice model on tensor product Hilbert space can be anomalous even when realized onsite. Here, onsite means that each Gauss law operator $W_p^{(g)}$ is expressed as a product of operators that act on onsite Hilbert spaces,
\begin{align}
    W_p^{(g)} = \bigotimes_{j\in \partial p} U^{(g)}_j~,
\end{align}
where $j$ labels the onsite Hilbert space $\mathcal{H}_j$ supported within a thin strip along $\partial p$. The onsite operator $U^{(g)}_j$ then satisfies the group algebra: $U^{(g)}_jU^{(g')}_j=U^{(gg')}_j$ with $g,g'\in G$. For instance, the $\Z_2$ 1-form symmetry in the (2+1)D $\Z_2$ toric code that corresponds to an emergent fermion $\psi$ is onsite in the above sense (i.e., a product of Pauli $X$ and $Z$ operators), while being anomalous. We will explicitly compute the anomaly index of this  $\Z_2$ symmetry in Sec.~\ref{subsec:Z2example}.

For onsite symmetries, the reduced symmetry operator $\Omega(\epsilon_{01},\epsilon_{12},g_{012})$ in \eqref{eq:defineOmega} becomes an overall phase, so each local factor $O_e$ in $\Omega=\prod_e O_e$ contributes only a phase. Nevertheless, because the phase $O_e$ depends locally on the 0-form variables $\epsilon_{01},\epsilon_{12}$, restricting $\Omega$ to a subsystem $I$ and performing dimensional reductions to define $A_l,\omega$ can yield non-trivial effects. This local dependence on $\epsilon$ allows the system to exhibit a non-vanishing ’t Hooft anomaly index despite the onsite realization of the 1-form symmetry. See Sec.~\ref{subsec:Z2example} for an explicit example.
This is in contrast to onsite 0-form symmetries: once $\Omega$ becomes a phase, any further restriction of operators or dimensional reduction is trivial. Consequently, an onsite 0-form symmetry must always be free of ’t Hooft anomalies. 

\subsubsection{Dynamical consequence of Else-Nayak type index}

Here we show that the Else-Nayak type index becomes trivial, $\omega=0$, if the 1-form $G$  symmetry preserves a short-range entangled (SRE) state $\ket{\Psi}$, i.e.,  $W_p\ket{\Psi}=\ket{\Psi}$ for all plaquettes $p$. Therefore, the nontrivial index $\omega\neq 0$ forbids a symmetric SRE state, and indeed defines an 't Hooft anomaly of a microscopic lattice model on a tensor product Hilbert space.

Without loss of generality, we assume that the SRE state $\ket{\Psi}$ is a product state: $\ket{\Psi}= \bigotimes_{j}\ket{0}_j$ where $j$ labels the onsite Hilbert space $\mathcal{H}_j$, and $\ket{0}_j$ is some state of $\mathcal{H}_j$. The following argument directly extends to generic SRE state; for a generic SRE state $\ket{\Psi} = V(\bigotimes_{j}\ket{0}_j)$ using a finite-depth circuit $V$, one can obtain a symmetry of the product state by conjugation $U'(\epsilon):=V^\dagger U(\epsilon) V$, and the index $\omega$ of $\ket{\Psi}$ reduces to that of a product state with this conjugated symmetry action.

Then, on each plaquette $p$ along $\partial R$, there exists a restricted Gauss law operator $W_{p;R}$ which preserves the product state $\ket{\Psi}$:
\begin{align}
    W_{p;R}\ket{\Psi} = \ket{\Psi}~.
    \label{eq:WpreservesSRE}
\end{align}
To see this, let us choose any restriction of $W_p$ to the region $R$, which we denote by $W'_{p;R}$.
Since the operator $W_p$ preserves $\ket{\Psi}$,  $W'_{p;R}$ acts on the product state $\ket{\Psi}$ by~\cite{li2024anyon}
\begin{align}
    W'_{p;R}\left(\bigotimes_{j}\ket{0}_j\right) = \ket{e}\otimes \ket{e'}\otimes \left(\bigotimes_{j\in\overline{e,e'}}\ket{0}_{j}\right)~,
\end{align}
where $e,e'$ are a pair of local regions at the intersection between $\partial R$ and $\partial p$, and $\ket{e}, \ket{e'}$ are local states. The rest of the Hilbert space has the product state. Then, one can use some local operators $V_e, V_{e'}$ at $e,e'$ to redefine a restriction $W_{p,R}= V_eV_{e'}W'_{p,R}$ so that \eqref{eq:WpreservesSRE} is satisfied.

This implies that $U_R(\epsilon)$ in \eqref{eq:explicit UR}, and hence $\Omega$, preserve the state $\ket{\Psi}$: $\Omega\ket{\Psi}=\ket{\Psi}$. 
Since $\Omega$ is a product of separate local operators $O_e$, each local operator $O_e$ preserves the state: $O_e\ket{\Psi}=\ket{\Psi}$.
This further implies that $F_e=A_l=0$, and hence $\omega=0$ on the state $\ket{\Psi}$.

\subsubsection{Graphical representation of Else-Nayak type index}
\label{sec:graphical}
\begin{figure}[tbh]
\centering
\includegraphics[width=0.6\textwidth]{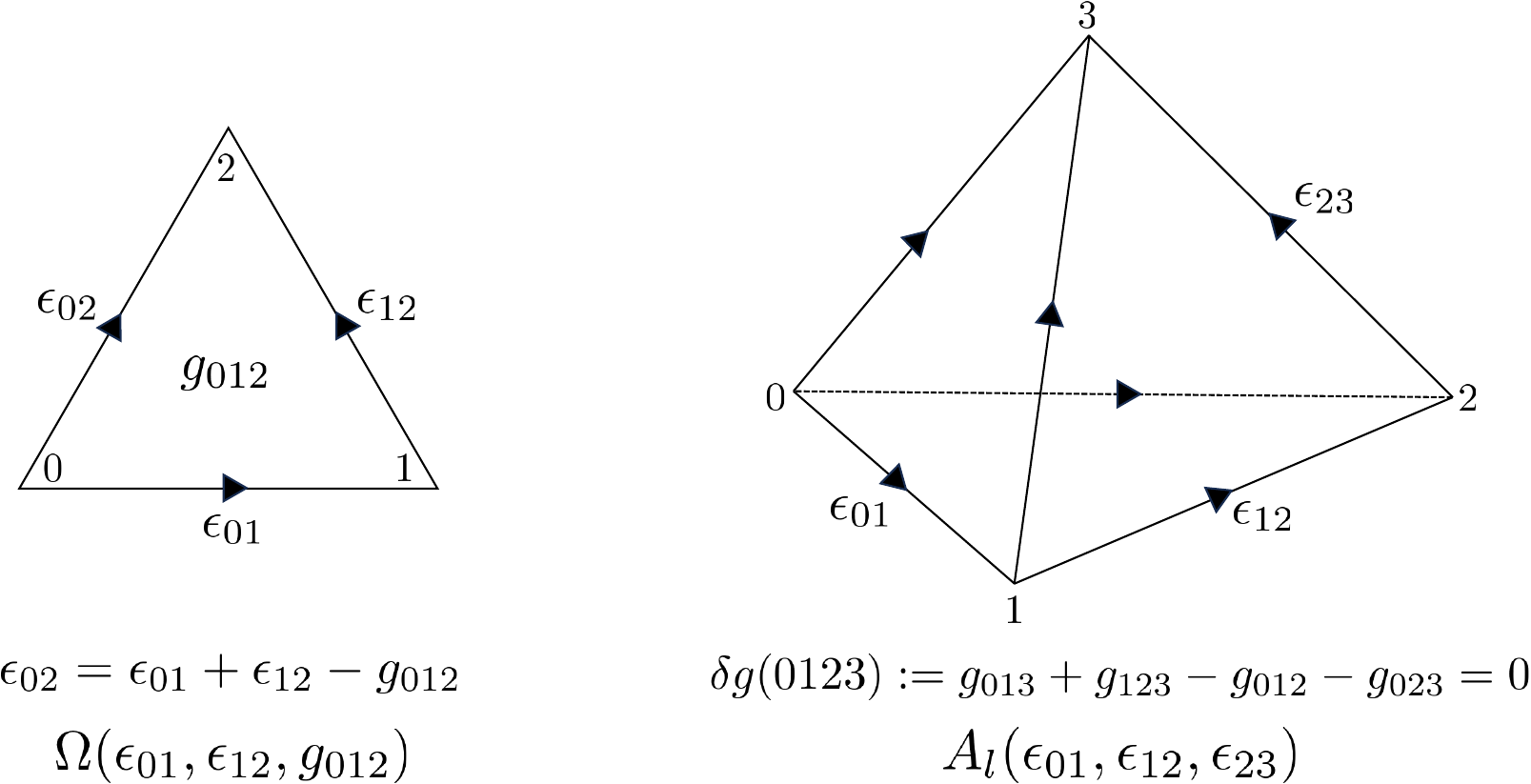}
\caption{The 0-forms $\epsilon_{ij}$, group elements $g_{ijk}$ are associated with the 1-simplices, 2-simplices of a simplicial complex. The operators $\Omega,A$, and $\omega$ are associated with 2,3,4-simplices respectively.}
\label{fig:omegagamma}
\end{figure}

The above procedure for getting the index $\omega$ is associated with a clear graphical interpretation. First, let us consider a certain simplicial complex where a 1-simplex $\langle ij\rangle$ is associated with  $\epsilon_{ij}\in C^0(\Lambda,G)$, and a 2-simplex $\langle ijk \rangle$ is associated with $g_{ijk}\in G$. On each 2-simplex we have the equation
\begin{align}
    \epsilon_{02} = \epsilon_{01} +\epsilon_{12}-g_{012}~,
\end{align}
as shown in Fig.~\ref{fig:omegagamma}.

One can then associate the operator $U(\epsilon_{01})$ with a 1-simplex $\langle 01 \rangle$ $\Omega(\epsilon_{01},\epsilon_{12}, g_{012})$ with a 2-simplex $\langle 012\rangle$, $A_l(\epsilon_{01},\epsilon_{12},\epsilon_{23},\{g\})$ with a 3-simplex $\langle 0123 \rangle$, and $\omega(\epsilon_{01},\epsilon_{12},\epsilon_{23},\epsilon_{34},\{g\})$ with a 4-simplex, see Fig.~\ref{fig:omegagamma}.
\footnote{This construction is inspired by the Kan complex~\cite{Kan1958} approach to classifying spaces for higher groups, as reviewed in Appendix~L of Ref.~\cite{Lan2019Fermiondecoration}.}

Now, the operations $\delta$ introduced in \eqref{eq:chi}, \eqref{eq:delta phi} are precisely the coboundary operation $\delta$ of cochains on this simplicial complex. The equations to obtain $\Omega, A,\omega$ in \eqref{eq:defineOmega}, \eqref{eq:Gamma}, \eqref{eq:omega def}, correspond to coboundary operations on the simplicial complex, in the sense that e.g., $\omega$ on a 4-simplex is obtained by a combination of $A_l$'s associated with the boundary 3-simplices.

\subsection{Example: $\Z_2$ toric code in (2+1)D}

As an explicit example, let us compute the anomaly index $[\omega_4]\in H^4(B^2G, U(1))$ for the $\Z_2$ 1-form symmetry of the (2+1)D $\Z_2$ toric code. This $\Z_2$ 1-form symmetry is onsite, given by a product of Pauli $X,Z$ operators along a string. 

\begin{figure}[tbh]
\centering
\includegraphics[width=0.4\textwidth]{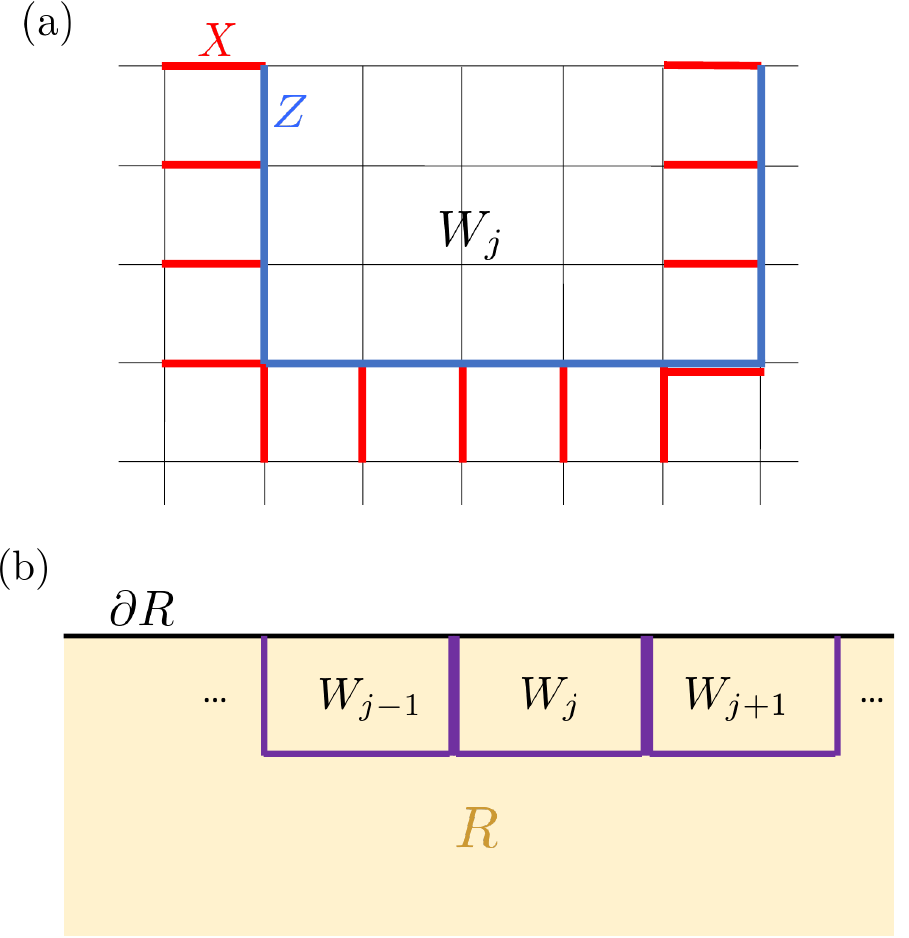}
\caption{{
(a): The truncated symmetry operator $W_j$ of the $\Z_2$ toric code. A single qubit is located at each edge of a square lattice, and the operator is a product of Pauli $X$ and $Z$ operators. (b): At the boundary of the region $R$, there is an array of truncated operators $\ldots, W_{j-1}, W_j, W_{j+1}, \ldots$ The neighboring operators anti-commute with each other due to fermionic statistics.}
}
\label{fig:toriccode_W}
\end{figure}

Let us label the vertices of $\Lambda$ at $\partial R$ by numbers $\ldots,j-1, j,j+1,\ldots$, so that each truncated Gauss law operator $W_{p;R}$ is simply denoted by $W_j$, where a plaquette $p$ corresponds to a vertex $j$ of $\Lambda$. See Fig.~\ref{fig:toriccode_W} for an illustration.
Since the 1-form symmetry is onsite, each truncated Gauss law operator $W_{j}$ again satisfies the $\Z_2$ algebra,
\begin{align}
    (W_{j})^2= 1~.
    \label{eq:W^2=1}
\end{align}
Also, reflecting that the string operator is a fermion, $W_{j}$ follows the commutation relation
\begin{align}
    W_{j}W_{j+1} = - W_{j+1}W_j~.
    \label{eq:Wcommutator}
\end{align}
Let us now compute the $H^4$ index of this $\Z_2$ 1-form symmetry. The truncated symmetry operator $W_R(\eps)$ has the form of
\begin{align}
    U_R(\epsilon) = \left(...W_{j-1}^{\epsilon(j-1)} W_{j}^{\epsilon(j)}W_{j+1}^{\epsilon(j+1)}...\right) \times \prod_{p\in\text{Int}(R)} (W_p)^{\eps(p)}~.
\end{align}
Using \eqref{eq:W^2=1}, \eqref{eq:Wcommutator}, the reduced operator $\Omega(\epsilon_{01},\epsilon_{12}, g_{012}) =U_R(\epsilon_{01})U_R(\epsilon_{12})    U_R(\epsilon_{01}+\epsilon_{12}-g_{012})^{-1}$ is expressed as
\begin{align}
    \Omega(\eps_{01},\eps_{12},g_{012}) = (-1)^{\sum_j (\eps_{01}(j)+\eps_{12}(j))g_{012} + \eps_{01}(j+1)\eps_{12}(j) } \times U_R(g_{012})~.
\end{align}
The contribution from $U_R(g_{012})$ provides only an overall factor in $F_e$ that is independent of the 0-cochains $\epsilon$ and hence does not affect the $H^4$ index.
We therefore focus on the remaining phase factor
\begin{align}
    O_e(\epsilon_{01},\epsilon_{12},g_{012})
    := (-1)^{\sum_j \bigl(\epsilon_{01}(j)+\epsilon_{12}(j)\bigr) g_{012}
    \;+\; \epsilon_{01}(j+1)\epsilon_{12}(j)}~,
\end{align}
which depends only on the local values of $\epsilon_{01}$ and $\epsilon_{12}$ near the edge $e=\langle j,j+1\rangle\subset\partial R$.
From this local functional $O_e$, we obtain
\begin{align}
    \begin{split}
        F_e(\eps_{01},\eps_{12},\eps_{23},\{g\}) &= \frac{1}{2}\left((\eps_{01}(j)+\eps_{12}(j))g_{012} + \eps_{01}(j+1)\eps_{12}(j)+(\eps_{02}(j)+\eps_{23}(j))g_{023} + \eps_{02}(j+1)\eps_{23}(j)\right) \\
        & - \frac{1}{2}\left((\eps_{12}(j)+\eps_{23}(j))g_{123} + \eps_{12}(j+1)\eps_{23}(j)+(\eps_{01}(j)+\eps_{13}(j))g_{013} + \eps_{01}(j+1)\eps_{13}(j)\right)~. \end{split}
\end{align}
By a repeated use of $\epsilon_{ik} = \eps_{ij}+\eps_{jk}-g_{ijk}$ and $g_{013}+g_{123}=g_{012}+g_{023}$ mod 2, the above expression of $F_e$ is rewritten as
\begin{align}
    F_e(\eps_{01},\eps_{12},\eps_{23},\{g\}) = \frac{1}{2}(\eps_{01}(j)~g_{123} +\eps_{01}(j+1)~g_{123}) +\phi(\{g\}) \quad \mod 1~,
\end{align}
where $\phi(g)$ is a functional that solely depends on $\{g\}$.
We again ignore the factor $\phi(\{g\})$ independent of 0-cochains $\epsilon$ since it does not affect the computation of the $H^4$ index. Then $A_j$ with $F=dA$ is given by
\begin{align}
    A_j = \frac{1}{2}\eps_{01}(j) ~g_{123}~.
\end{align}
Finally, the index $\omega_4$ is given by
\begin{align}
    \begin{split}
\omega_l(\epsilon_{01},\epsilon_{12},\epsilon_{23},\epsilon_{34},\{g\}) &=A_j(\epsilon_{01},\epsilon_{12},\epsilon_{23}) + A_j(\epsilon_{01},\epsilon_{13},\epsilon_{34}) + A_j(\epsilon_{12},\epsilon_{23},\epsilon_{34})  
- A_j(\epsilon_{02},\epsilon_{23},\epsilon_{34}) - A_j(\epsilon_{01},\epsilon_{12},\epsilon_{24})~\\
&= \frac{1}{2}\eps_{01}~(g_{123}+g_{134}-g_{124})+\frac{1}{2}\eps_{12}~g_{234}-\frac{1}{2}\eps_{02}~g_{234} \\
&= \frac{1}{2}g_{012}~g_{234}~.
\end{split}
\end{align}
Therefore we get
\begin{align}
    \omega_4(\{g\}) = \frac{1}{2}g_{012}~g_{234}~,
\label{eq: omega_4 SM}
\end{align}
which can be written as $\frac{1}{2}g\cup g$ evaluated at a 4-simplex $\langle 01234\rangle$, by regarding $g$ as a 2-cocycle $g\in Z^2(B^2G,\Z_2)$. 
Here, $\cup$ is a cup product of cochains on the triangulation $\Lambda$, defined as
\begin{align}
    a_k\cup a_l(0,\dots, k+l) = a_k(0,\dots, k)a_l(k,\dots, l)~,
\end{align}
with $k,l$-cochains $a_k,a_l$. The cup product generally induces a map between cohomology classes, $\cup: H^k\times  H^l\to H^{k+l}$.
Explicitly, $g$ is a $\mathbb{Z}_2$-valued $2$-cocycle taking values $0$ or $1$; consequently, $\omega_4$ takes values $0$ or $\tfrac12$ in Eq.~\eqref{eq: omega_4 SM}, where we embed $\Z_2$ in $U(1)=\mathbb{R}/\mathbb{Z}$.
Therefore $\omega$ defines an element of $Z^4(B^2G,U(1))$, producing the desired (3+1)D response action for the anomaly~\cite{Chen:2017fvr, Chen:2018nog, Chen2020}.

\section{Higher-form symmetry and anomaly}
\label{sec:higher}
In this section, we describe the anomaly index of finite higher-form symmetry in generic spacetime dimensions. While it is not straightforward to extend the Else-Nayak type index for generic circuits in spacetime dimensions higher than (2+1)D, we describe generalizations by assuming a specific form of tensor product Hilbert space and symmetry actions. Given a finite abelian group $G$, we consider a triangulation $\Lambda$ in the $d$-dimensional space, and assume that a Hilbert space is given by tensor product of an $|R|$-dimensional local Hilbert space on each $p$-simplex of $\Lambda$, with $R$ being a finite $G$-module. For each local Hilbert space, the basis state is labeled by $\ket{r}$ with $r\in R$. Therefore the basis state of the whole Hilbert space is labeled by $\ket{a}$, with $a\in C^p(\Lambda,R)$ a degree $p$ cochain of $\Lambda$. We note that unlike the setup of Sec.~\ref{sec:1form}, now the triangulation $\Lambda$ is not mesoscopic; a single onsite Hilbert space lives on each $p$-simplex of the triangulation.

We also make an assumption on the action of $p$-form $G$ symmetry on the Hilbert space. 
A symmetry operator is labeled by a $(p-1)$-cochain $\epsilon\in C^{p-1}(\Lambda,G)$. Then we assume that the symmetry acts by
\begin{align}
    U(\epsilon)\ket{a} = e^{2\pi i \int F[a,\epsilon]}\ket{a+d\epsilon}~,
    \label{eq: symmetry action in generic dim}
\end{align}
with a $d$-form $F[a,\epsilon]\in C^{d}(\Lambda,G)$ which is a local functional of $a,\epsilon$. The action of $g\in G$ on $r\in R$ is denoted as $r+g$ by an abuse of notation. We further assume that $U(\epsilon)$ satisfies the property
\begin{equation}
    U(\eps_1+\eps_2)=U(\eps_1)U(\eps_2)~,
    \label{eq:homomorphism in generic dim}
\end{equation}
and also
\begin{align}
    U(d\eta) = 1~,
    \label{eq:topological in generic dim}
\end{align}
with a $(p-2)$-cochain $\eta\in C^{p-2}(\Lambda,G)$. The property $\eqref{eq:topological in generic dim}$ is not well-defined for $1$-form symmetry, $p=1$. In the case of $p=1$, we instead assume the property 
\begin{align}
    U(g) = 1~,
    \label{eq:topological for p=1}
\end{align}
with a constant 0-form $g\in C^0(\Lambda, G)$ labeled by $g\in G$.

The Gauss law operator $W_\Delta$ is supported at a single $(p-1)$-simplex $\Delta$, and corresponds to the operator $U(\epsilon_\Delta)$ with the $(p-1)$-cochain $\epsilon_\Delta$ where $\epsilon_\Delta$ takes the nonzero value $g\in G$ on a single $(p-1)$-simplex $\Delta$, otherwise zero. Note that the above property \eqref{eq:homomorphism in generic dim} is equivalent to commutativity of Gauss laws: $[U(\epsilon_\Delta),U(\epsilon_{\Delta'})]=1$. Also, note that \eqref{eq:topological in generic dim}, \eqref{eq:topological for p=1} is equivalent to requiring that the symmetry operators become topological when the Gauss law $W_\Delta=1$ is enforced. Therefore, the above properties \eqref{eq:homomorphism in generic dim}, \eqref{eq:topological in generic dim}, \eqref{eq:topological for p=1} are necessary and sufficient for qualifying $U(\epsilon)$ as a $p$-form symmetry.

We note that this assumption on the symmetry action is a natural generalization of the 0-form symmetry action assumed in \cite{ElseNayak2014} in (2+1)D and higher, which takes $p=0$ with the symmetry action
\begin{align}
    U(g)\ket{a} = e^{2\pi i \int F[a,g]}\ket{a+g}~,
\end{align}
with a 0-form $a\in C^0(\Lambda,R)$ and $g\in G$.
In Ref.~\cite{ElseNayak2014}, the anomaly index of 0-form symmetry is derived for this specific symmetry action through the dimension reduction of symmetry operators, which we will generalize to higher-form symmetries below.

\subsection{Else-Nayak type reduction for higher-form symmetries}

\subsubsection{Warm-up: 1-form symmetry}
\label{subsubsec:1-form symmetry in generic dim}
\begin{figure}[b]
\centering
\includegraphics[width=0.6\textwidth]{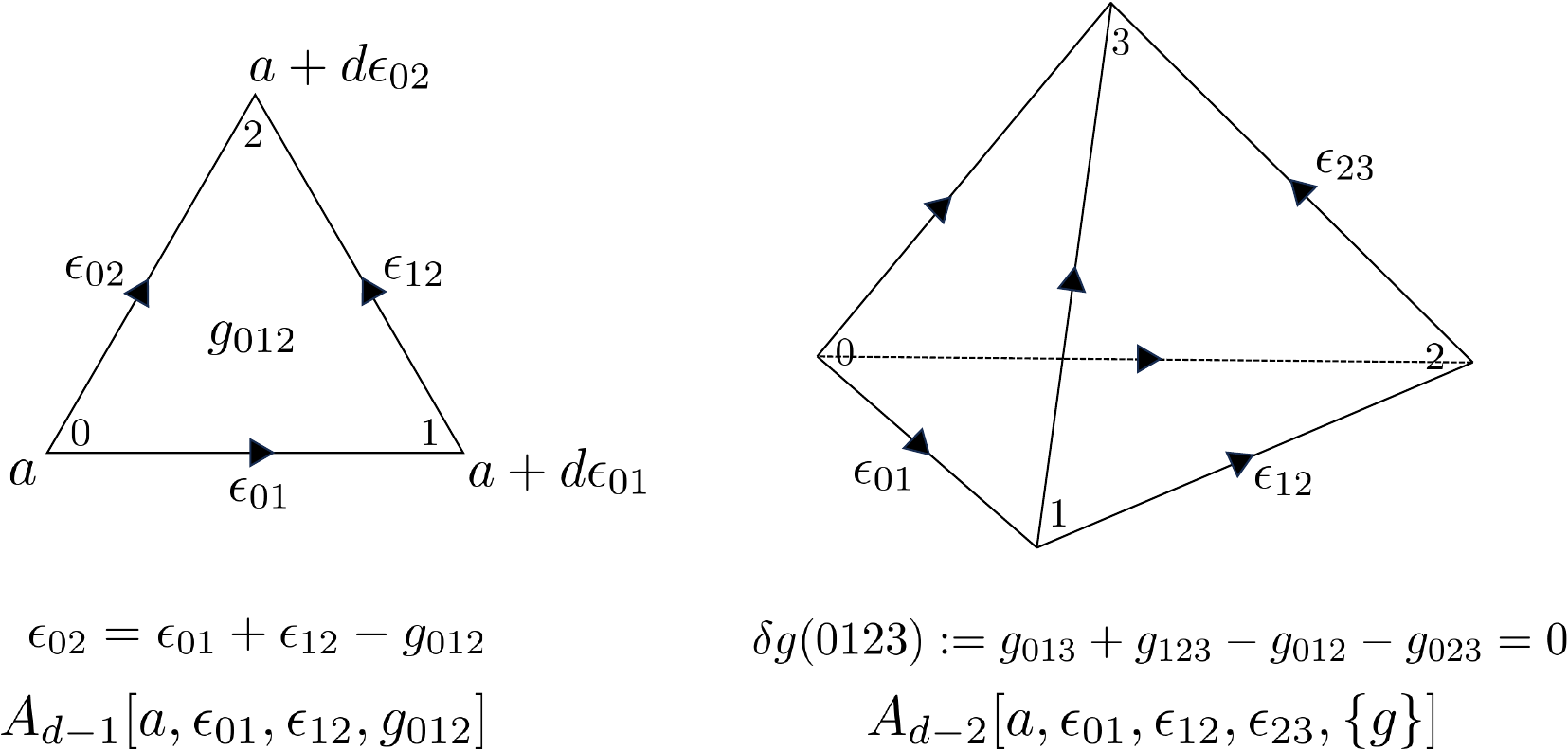}
\caption{The 0-forms $\epsilon_{ij}$, group elements $g_{ijk}$ are associated with the 1-simplices, 2-simplices of a simplicial complex. The state label $a$ is associated with a 0-simplex. The $j$-form $A_j$ is associated with a $(d+1-j)$-simplex.}
\label{fig:Aj}
\end{figure}

Let us begin with the 1-form $G$  symmetry in generic $d$ spatial dimensions, with $p=1$.
Suppose that there is a Hilbert space with a base labeled by 1-form configurations $a$, with $p=1$. The anomalous symmetry is generated by $U(\epsilon)$ introduced in \eqref{eq: symmetry action in generic dim}. In this case with $p=1$, \eqref{eq:topological in generic dim} is replaced by the property
\begin{align}
    U(g) = 1~,
\end{align}
with a constant 0-form labeled by $g\in G$. This implies that the product of Gauss law operators $W_\Delta$ over the 0-simplices in the whole space cancels out, ensuring that the 1-form symmetry operator becomes topological by enforcing the Gauss law $W_\Delta=1$. 
Since $U$ is a homomorphism \eqref{eq: symmetry action in generic dim}, $U(\epsilon)$ satisfies
\begin{equation}
    U(\eps+g)=U(\eps)~.
\end{equation}
Therefore, by combining with \eqref{eq: symmetry action in generic dim} we obtain a general relation satisfied by the functional $F$, that is 
\begin{equation}
    \int F[a,\eps_{01}]+F[a+d\eps_{01},\eps_{12}]-F[a,\eps_{01}+\eps_{12}-g_{012}]=0~,
    \label{eq:deltaF}
\end{equation}
with $\epsilon_1,\epsilon_2\in C^0(\Lambda,G)$, $g_{012}\in G$.
We graphically represent this by associating $F[a,\eps]$ with the line labeled by $\eps$, and place a $g$ label on the interior of each triangle loop with three lines $\eps_{01},\eps_{12},\eps_{02}$, satisfying
\begin{equation}
    \eps_{01}+\eps_{12}-\eps_{02}=g_{012}~,
\end{equation}
on a 2-simplex $(012)$. See Fig.~\ref{fig:Aj}.
We use an expression $D\eps=g$ to summarize the above relationship for a 2-simplex. 

The equation \eqref{eq:deltaF} implies that the combination $F[a,\eps_{01}]+F[a+d\eps_{01},\eps_{12}]-F[a,\eps_{01}+\eps_{12}-g_{012}]$ is exact. Therefore there exists a $(d-1)$-form $A_{d-1}\in C^{d-1}(\Lambda,G)$ satisfying
\begin{equation}
    dA_{d-1}[a,\eps_{01},\eps_{12},g_{012}]=F[a,\eps_{01}]+F[a+d\eps_{01},\eps_{12}]-F[a,\eps_{01}+\eps_{12}-g_{012}]~.
    \label{eq: reduction to AI}
\end{equation}
This allows us to perform ``dimensional reduction'' of the symmetry operator. Let us take a $d$-dimensional disk $D^d$ in the triangulation $\Lambda$, then we define a restriction of the symmetry operator within $D^d$ by
\begin{align}
    U_{D^d}(\epsilon)\ket{a} = e^{2\pi i \int_{D^d} F[a,\epsilon]}\ket{a+d\epsilon}~.
\end{align}
According to \eqref{eq: reduction to AI}, $U_{D^d}(\epsilon)$ satisfies
\begin{align}
\Omega_{\partial D^d}(\epsilon_{01},\epsilon_{12},g_{012}) = U_{D^d}(\epsilon_{01}+\epsilon_{12}-g_{012})^{-1}U_{D^d}(\epsilon_{12})U_{D^d}(\epsilon_{01})~,
\end{align}
with $\Omega_{\partial D^d}=\Omega_{S^{d-1}}$ an operator supported along $\partial D^d$, defined by
\begin{align}
    \Omega_{S^{d-1}}(\epsilon_{01},\epsilon_{12},g_{012})\ket{a} =  e^{2\pi i\int_{S^{d-1}}A_{d-1}[a,\eps_{01},\eps_{12},g_{012}]}\ket{a}~.
\end{align}

Such a dimensional reduction process can be performed iteratively until the degree of a cochain $A_j$ reaches $j=-1$. Every reduction process has the form of $dA_{j-1}=\delta A_j$, where $\delta$ is a coboundary operation of $A_j[\{\epsilon\},\{g\}]$ introduced below. Let us explicitly write down the first few steps of the reduction: 
\begin{equation}
    dA_{d-1}[a,\eps_{01},\eps_{12},g_{012}]=F[a,\eps_{01}]+F[a+d\eps_{01},\eps_{12}]-F[a,\eps_{01}+\eps_{12}-g_{012}]~,
    \label{eq:Ad-1}
\end{equation}
\begin{align}
    dA_{d-2}[a,\eps_{01},\eps_{12},\eps_{23},g_{012},g_{013},g_{023},g_{123}]&=A_{d-1}[a+d\eps_{01},\eps_{12},\eps_{23},g_{123}]-A_{d-1}[a,\eps_{01}+\eps_{12}-g_{012},\eps_{23},g_{023}]\notag\\&+A_{d-1}[a,\eps_{01},\eps_{12}+\eps_{23}-g_{123},g_{013}]-A_{d-1}[a,\eps_{01},\eps_{12},g_{012}]~,
    \label{eq:Ad-2}
\end{align}
\begin{align}
    dA_{d-3}[a,\eps_{01},\cdots,\eps_{34},g_{012},\cdots,g_{234}]&=A_{d-2}[a+d\eps_{01},\eps_{12},\eps_{23},\eps_{34},g_{123},g_{124},g_{134},g_{234}]\notag \\
    &-A_{d-2}[a,\eps_{01}+\eps_{12}-g_{012},\eps_{23},\eps_{34},g_{023},g_{024},g_{034},g_{234}]\notag\\
    &+A_{d-2}[a,\eps_{01},\eps_{12}+\eps_{23}-g_{123},\eps_{34},g_{013},g_{014},g_{034},g_{134}]\notag\\
    &-A_{d-2}[a,\eps_{01},\eps_{12},\eps_{23}+\eps_{34}-g_{234},g_{012},g_{014},g_{024},g_{124}]\notag \\ 
    &+A_{d-2}[a,\eps_{01},\eps_{12},\eps_{23},g_{012},g_{013},g_{023},g_{123}]~.
    \label{eq:Ad-3}
\end{align}

These operations $\delta$ admit a clear geometric interpretation as a coboundary of cochains on a simplicial complex. To see this, we again introduce a simplicial complex, where a 1-simplex $\langle ij\rangle$ is associated with  $\epsilon_{ij}\in C^1(\Lambda,G)$, and a 2-simplex $\langle ijk \rangle$ is associated with $g_{ijk}\in G$. On each 2-simplex we have the equation
\begin{align}
    \epsilon_{02} = \epsilon_{01} +\epsilon_{12}-g_{012}~,
\end{align}
as shown in Fig.~\ref{fig:Aj}. We associate a state label $\ket{a}$ with a 0-simplex, and $F[a,\epsilon_{01}]$ as a 1-simplex $\langle 01\rangle$ where the 0-simplex $0$ has $\ket{a}$. One can then see that each functional $A_{j}$ introduced above is associated with $(d+1-j)$-simplex, e.g., $A_{d-1}[a,\eps_{01},\eps_{12},g_{012}]$ is at a 2-simplex $\langle 012\rangle$. Now, the above operations $\delta$ are precisely the coboundary operations evaluated at a single simplex. For instance, the rhs of \eqref{eq:Ad-3} is a sum over $A_{d-2}$ on boundary 3-simplices of a single 4-simplex $\langle 01234\rangle$, and evaluates the coboundary $\delta A_{d-2}$ of a ``3-cochain'' $A_{d-2}$ evaluated at a 4-simplex $\langle 01234\rangle$.

Associated with the above dimensional reduction process, there is an iterative process for the dimensional reduction of symmetry operators. That is, the $j$-form $A_j$ is associated with a $j$-dimensional operator $\Omega_{S^j}$ supported at a $j$-sphere embedded in the triangulation $\Lambda$:
\begin{align}
    \Omega_{S^j}\ket{a} = e^{2\pi i \int_{S^j} A_j} \ket{a}~.
\end{align}
Then, take a restriction of the operator $\Omega_{S^j}\ket{a}$ to a $j$-dimensional hemisphere (disk) $D^j$ to define an operator $\Omega_{D^j}$.
Then, the reduction equation $dA_{j-1}=\delta A_j$ implies that the $(j-1)$-dimensional operator $\Omega_{S^{j-1}}=\Omega_{\partial D^{j}}$ is obtained from a product of $\Omega_{D^j}$. For instance, the equations \eqref{eq:Ad-1}, \eqref{eq:Ad-2}, \eqref{eq:Ad-3} lead to
\begin{align}
    \Omega_{\partial D^d}(\epsilon_{01},\epsilon_{12},g_{012}) = U_{D^d}(\epsilon_{01}+\epsilon_{12}-g_{012})^{-1}U_{D^d}(\epsilon_{12})U_{D^d}(\epsilon_{01})~,
\end{align}
\begin{align}
    \Omega_{\partial D^{d-1}}(\eps_{01},\eps_{12},\eps_{23},g_{012},\dots,g_{123}) &= U(\epsilon_{01})^{-1}\Omega_{D^{d-1}}(\eps_{12},\eps_{23},g_{123})U(\epsilon_{01})\Omega_{D^{d-1}}(\eps_{01}+\eps_{12}-g_{012},\eps_{23},g_{023})^{-1}\notag \\
    &\times \Omega_{D^{d-1}}(\eps_{01},\eps_{12}+\eps_{23}-g_{123},g_{013})\Omega_{D^{d-1}}(\eps_{01},\eps_{12},g_{012})^{-1}~,
\end{align}
\begin{align}
    \Omega_{\partial D^{d-3}}(a,\eps_{01},\cdots,\eps_{34},g_{012},\cdots,g_{234})&=U(\epsilon_{01})^{-1}\Omega_{D^{d-2}}(\eps_{12},\eps_{23},\eps_{34},g_{123},g_{124},g_{134},g_{234})U(\epsilon_{01}) \notag \\
    &\times\Omega_{D^{d-2}}(\eps_{01}+\eps_{12}-g_{012},\eps_{23},\eps_{34},g_{023},g_{024},g_{034},g_{234})^{-1}\notag\\
    &\times\Omega_{D^{d-2}}(\eps_{01},\eps_{12}+\eps_{23}-g_{123},\eps_{34},g_{013},g_{014},g_{034},g_{134})\notag\\
    &\times \Omega_{D^{d-2}}(\eps_{01},\eps_{12},\eps_{23}+\eps_{34}-g_{234},g_{012},g_{014},g_{024},g_{124})^{-1}\notag \\ 
    &\times \Omega_{D^{d-2}}(\eps_{01},\eps_{12},\eps_{23},g_{012},g_{013},g_{023},g_{123})~.
    \label{eq:Omega d-3}
\end{align}
We note that when $d=2$, the expression \eqref{eq:Omega d-3} of $\Omega_{\partial D^{d-3}}$ coincides with the form of the anomaly index \eqref{eq:omega def} for generic 1-form symmetry in (2+1)D, where $\Omega_{\partial D^{d-3}}$ corresponds to $e^{2\pi i \omega}$, and $\Omega_{\partial D^{d-2}}$ to $e^{2\pi i A_l}$. 

\subsubsection{Example: Anomalous $\Z_2$ 1-form symmetry in (2+1)D}
\label{subsec:Z2example}
Consider the anomalous $\Z_2$ 1-form symmetry in (2+1)D characterized by the (3+1)D response~\cite{Chen:2017fvr, Chen:2018nog, Chen2020}
\begin{equation}
    \pi i \int B\cup B~,
    \label{eq:BcupB}
\end{equation}
with $B$ the 2-form $\Z_2$ background gauge field.

Let us also take $R=\Z_2$, so that the state $\ket{a}$ is labeled by $a\in C^1(\Lambda,\Z_2)$.
The anomalous symmetry is realized by choosing a functional $F$ as
\begin{equation}
    F[a,\eps]=\frac{1}{2}a\cup d\eps,
\end{equation}
where $\cup$ is a cup product of cochains on the triangulation $\Lambda$, defined as
\begin{align}
    a_k\cup a_l(0,\dots, k+l) = a_k(0,\dots, k)a_l(k,\dots, l)~,
\end{align}
with $k,l$-cochains $a_k,a_l$.

\begin{figure}[tbh]
\centering
\includegraphics[width=0.15\textwidth]{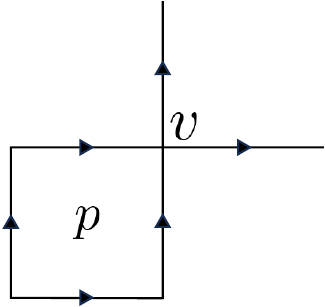}
\caption{The configurations of a vertex and plaquette to define a symmetry operator $U(\hat{v})$. The arrows represent directions of edges used to define the cup product (associated with a branching structure of the triangulation).}
\label{fig:plaquette}
\end{figure}

Since $R=\Z_2$, the Hilbert space is regarded as a qubit system with a single qubit on each edge of the triangulation. A Pauli $Z$ operator is associated with the value of $a$: $Z_e\ket{a} = (-1)^{a(e)}\ket{a}$ on each edge $e$.
With the above choice of $F[a,\eps]$, the symmetry operator $U(\epsilon)$ has the expression of $X$-star terms coupled to $Z$-plaquette terms. For instance, let us consider a square lattice (each square consists of two simplices of $\Lambda$). If $\epsilon = \hat{v}$, where $\hat{v}=1$ on a single vertex $v$ and zero otherwise, $U(\epsilon)$ is given by
\begin{align}
    U(\hat{v}) = \left(\prod_{e\in \partial p} Z_e\right) \left(\prod_{v\in \partial e} X_e\right)~,
\end{align}
where $p$ is a single plaquette shown in Fig~\ref{fig:plaquette}.
This is a Gauss law operator which corresponds to a small closed $\psi$-string operator of the (2+1)D $\Z_2$ toric code.

Now we perform the reduction above, and after calculation we have a rather simple expression
\begin{align}
    A_1[a,\eps_{01},\eps_{12},g_{012}]&=\frac{1}{2}\eps_{01}\cup d\eps_{12}~,\\
    A_{0}[a,\eps_{01},\eps_{12},\eps_{23},g_{012},g_{013},g_{023},g_{123}]&=\frac{1}{2}g_{012}\eps_{23}~,\\
    A_{-1}[a,\eps_{01},\cdots,\eps_{34},g_{012},\cdots,g_{234}]&=\frac{1}{2}g_{012}g_{234}~.
\end{align}
One can see that $A_{-1}$ can be written as $\frac{1}{2}g\cup g$ evaluated at a 4-simplex $\langle 01234\rangle$, by regarding $g$ as a 2-cocycle $g\in Z^2(B^2G,U(1))$ satisfying $\delta g=1$. Therefore $\omega$ defines an element of $Z^4(B^2G,U(1))$, and produces the desired (3+1)D response action for the 't Hooft anomaly \eqref{eq:BcupB}.
Below we will generally show that $\omega$ defines a representative of $H^{d+2}(B^{p+1}G,U(1))$ for $p$-form $G$ symmetry in $(d+1)$ spacetime dimensions.

\subsubsection{Higher-form symmetry}\label{Section:Higher form symmetry}

We then illustrate the reduction process for higher-form symmetries in generic $(d+1)$ spacetime dimensions. We start by placing $p$-form configuration $a$ on each point, and $(p-1)$-form labels $\eps$ on 1-cells. In general, we place a $(p-k)$-form label $\eta^{(k)}$ on each $k$-cell with $0\le k \le p+1$, and they satisfy
\begin{equation}
    D\eta^{(n)}=d\eta^{(n+1)}~,
\end{equation}
with $D$ representing the oriented sum of all simplicial faces on the boundary of a simplex, and $d$ is the ordinary differential in our base (spatial) manifold. When $k=p+1$, the label $\eta^{p+1}$ corresponds to an element $g\in G$. 

Suppose that the symmetry acts by operators $U(\eps)$ having the form of equation \eqref{eq: symmetry action in generic dim}. We could combine the properties \eqref{eq:homomorphism in generic dim} and \eqref{eq:topological in generic dim} together and obtain a generalization of the relation \eqref{eq:deltaF}, that is 

\begin{equation}
    \int F[a,\eta^{(1)}_{01}]+F[a+d\eta^{(1)}_{01},\eta^{(1)}_{12}]-F[a,\eta^{(1)}_{01}+\eta^{(1)}_{12}-d\eta^{(2)}_{012}]=0~,
\end{equation}
with $\eta^{(1)}_{01}, \eta^{(1)}_{12}\in C^{p-1}(\Lambda,G)$ and $\eta^{(2)}_{012}\in C^{p-2}(\Lambda,G)$. Therefore there exists a $(d-1)$-form $A_{d-1}\in C^{d-1}(\Lambda,G)$ satisfying

\begin{equation}
    dA_{d-1}[a,\eta^{(1)}_{01},\eta^{(1)}_{12},\eta^{(2)}_{012}]=F[a,\eta^{(1)}_{01}]+F[a+d\eta^{(1)}_{01},\eta^{(1)}_{12}]-F[a,\eta^{(1)}_{01}+\eta^{(1)}_{12}-d\eta^{(2)}_{012}]~.\label{reduction p1}
\end{equation}
Graphically, the function $A_{d-1}$ is defined on each labeled 2-cell, and $F$ is defined on each labeled 1-cell. The relation \eqref{reduction p1} lets us reduce the dimension of the base manifold by one. Such dimensional reduction process can be performed iteratively until the degree of the cochain reaches $-1$. For each $j>0$, there is a cochain $A_j$ defined on labeled $(j+1)$-cells, and every reduction process has the form of $dA_{j-1}=\delta A_j$, where $\delta$ is the coboundary operator dual to the boundary operator of labeled cells. 

The dimensional reduction process stops when the degree becomes $-1$, and $A_{-1}$ is just a phase in $U(1)$. As the base manifold becomes empty, the function $A_{-1}$ only depends on $g\in G$ labels on $(p+1)$-faces, which satisfy $Dg=0$. This means $A_{-1}$ is an object in $C^{d+2}(B^{p+1}G,U(1))$ (see Appendix \ref{app:EMspace} for a review of the Eilenberg-MacLane space $B^{p+1}G$ and its cohomology). From the last step of reduction, we obtain $A_{-1}=\delta A_0$, and therefore
\begin{equation}
    \delta A_{-1}=0~. 
\end{equation}
When we take redefinitions of $A_k$ during the reduction process, the cocycle $A_{-1}$ is only shifted by a coboundary $A_{-1}\mapsto A_{-1}+\delta \phi$. 
Therefore, the final result $A_{-1}$ gives a well-defined cohomology class in $H^{d+2}(B^{p+1}G,U(1))$. 

\section{Comments on statistical invariant}
\label{sec:statistics}

Aside from the Else-Nayak index, there is another way to define an anomaly index of 1-form symmetry in (2+1)D, as we will see below. For simplicity, let us work on $G=\Z_N$.

\begin{figure}[tbh]
\centering
\includegraphics[width=0.9\textwidth]{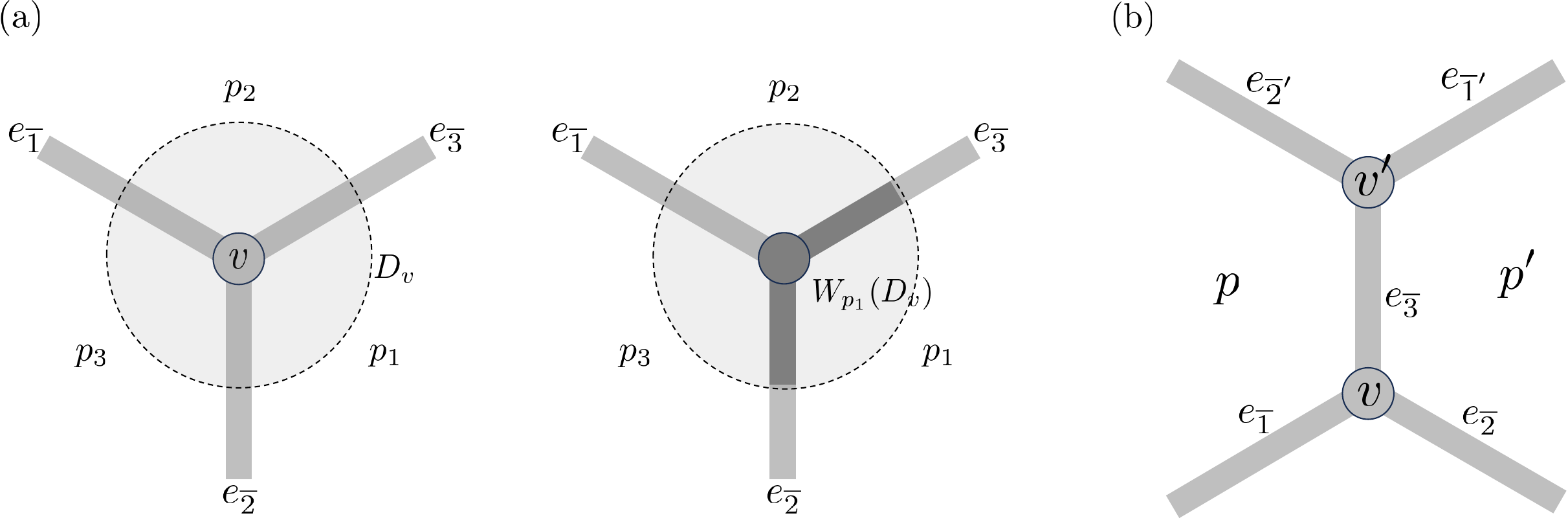}
\caption{(a) Left: the disk region $D_v$ around a vertex $v$ of the dual lattice. (a) Right: $W_p(D_v)$ is a truncation of $W_p$ within a region $D_v$. (b): The adjacent plaquettes $p,p'$ and edges nearby $p,p'$.}
\label{fig:Dv}
\end{figure}

Let us again consider a setup of Sec.~\ref{sec:1form} in (2+1)D, where the symmetry operators are supported at the mesoscopic dual lattice $\hat\Lambda$ as shown in Fig.~\ref{fig:duallattice}. 
Each vertex $v$ of the dual lattice $\hat{\Lambda}$ has three plaquettes and three edges adjacent to it. Let us denote the plaquettes $p_1, p_2, p_3$ (labeled anticlockwise), and edges $e_{\overline{1}}, e_{\overline{2}}, e_{\overline{3}}$ (an edge $\overline{j}$ is adjacent to the plaquettes with numbers different from $j$.)
We consider a disk $D_v$ enclosing the vertex $v$, whose boundary $\partial D_v$ cuts the three edges $e_{\overline{1}}, e_{\overline{2}}, e_{\overline{3}}$. See Fig.~\ref{fig:Dv} (a).

Then, we consider a truncation of the circuit $W_p$ to $D_v$, which we denote by $W'_p(D_v)$.
Due to \eqref{eq: product of Gauss laws}, the operators
\begin{align}
    W'_{p_2}(D_v)W'_{p_3}(D_v)W'_{p_1}(D_v)~, \quad W'_{p_3}(D_v)W'_{p_2}(D_v)W'_{p_1}(D_v)
\end{align}
are products of three local operators supported nearby the intersection between $\partial D_v$ and edges $e_{\overline{1}}, e_{\overline{2}}, e_{\overline{3}}$. For instance,
\begin{align}
    W'_{p_2}(D_v)W'_{p_3}(D_v)W'_{p_1}(D_v) = O_{\overline{1}} O_{\overline{2}} O_{\overline{3}}~,
\end{align}
where $O_{\overline{j}}$ is supported at the intersection $e_{\overline{j}}\cap \partial D_v$. Let us then redefine the truncated operators $W'$ as
\begin{align}
    W_{p_1}(D_v) = W'_{p_1}(D_v) O^\dagger_{\overline{2}} O^\dagger_{\overline{3}}~, \quad W_{p_3}(D_v) = W'_{p_3}(D_v) O^\dagger_{\overline{1}}~, \quad W_{p_2}(D_v) = W'_{p_2}(D_v)~.
\end{align}
This eliminates the local operators at $\partial D_v$, and we have
\begin{align}
    W_{p_2}(D_v)W_{p_3}(D_v)W_{p_1}(D_v) = 1~.
\end{align}
With these choices of truncation, $W_{p_3}(D_v)W_{p_2}(D_v)W_{p_1}(D_v)$ also becomes a trivial operator up to overall phase:
\begin{align}
    W_{p_3}(D_v)W_{p_2}(D_v)W_{p_1}(D_v) = e^{i\Theta_v}~.
\end{align}
This defines an invariant $\Theta_v\in U(1)$ on each vertex of the dual lattice.

Due to $W_{p_2}(D_v)W_{p_3}(D_v)W_{p_1}(D_v) = 1$, the circuits $W_p(D_v)$ can be expressed as
\begin{align}
    W_{p_1}(D_v) = U^\dagger_{e_{\overline{2}}} U_{e_{\overline{3}}}~, \quad W_{p_3}(D_v) = U^\dagger_{e_{\overline{1}}} U_{e_{\overline{2}}}~, \quad W_{p_2}(D_v) = U^\dagger_{e_{\overline{3}}} U_{e_{\overline{1}}}~,
\end{align}
with some finite-depth circuit $U_{e_{\overline{j}}}$ supported at $e_{\overline{j}}$. 
With this expression, the invariant is given by
\begin{align}
    U^\dagger_{e_{\overline{1}}} U_{e_{\overline{2}}}U^\dagger_{e_{\overline{3}}} U_{e_{\overline{1}}}U^\dagger_{e_{\overline{2}}} U_{e_{\overline{3}}} = e^{i\Theta_v}~,
\end{align}
which is the well-known T-junction invariant~\cite{Levin2003Fermions}.

The invariant $\Theta_v$ is independent of a vertex $v$.
This can be seen by evaluating the commutator $[W_p, W_{p'}]$ for a neighboring pair of plaquettes $p,p'$. Suppose that $p,p'$ share an edge $e_{\overline{3}}=\langle vv'\rangle$, and a vertex $v$ has three edges $e_{\overline{1}},e_{\overline{2}},e_{\overline{3}}$, while $v'$ has $e_{\overline{1}'},e_{\overline{2}'},e_{\overline{3}}$. See Fig.~\ref{fig:Dv} (b). Then, each circuit $W_p,W_{p'}$ has an expression
\begin{align}
    W_p = V U_{\overline{2}'}U_{\overline{3}}^\dagger U_{\overline{1}}~, \quad W_{p'} = V' U^\dagger_{\overline{2}}U_{\overline{3}} U^\dagger_{\overline{1}'}~,
\end{align}
where $V,V'$ are away from the edge $e_{\overline{3}}$, and $[V,U^\dagger_{\overline{1}'}]=[V,U_{\overline{2}}^\dagger]=[V,U_{\overline{3}}]=[V',U_{\overline{1}}]=[V',U_{\overline{2}'}]=[V',U_{\overline{3}}]=1$. Due to  $[W_p,W_{p'}]=1$, we have
\begin{align}
    U_{\overline{2}'}U_{\overline{3}}^\dagger U_{\overline{1}}U^\dagger_{\overline{2}}U_{\overline{3}} U^\dagger_{\overline{1}'} = U^\dagger_{\overline{2}}U_{\overline{3}} U^\dagger_{\overline{1}'}U_{\overline{2}'}U_{\overline{3}}^\dagger U_{\overline{1}}~.
\end{align}
Since $U^\dagger_{\overline{1}}U_{\overline{2}}U_{\overline{3}}^\dagger U_{\overline{1}}U^\dagger_{\overline{2}}U_{\overline{3}} = e^{i\Theta_v}$ and $[U_{\overline{1}},U^\dagger_{\overline{1}'}]=[U^\dagger_{\overline{2}},U_{\overline{2}'}]=1$, the lhs is rewritten as
\begin{align}
    U_{\overline{2}'}(U_{\overline{3}}^\dagger U_{\overline{1}}U^\dagger_{\overline{2}}U_{\overline{3}})U^\dagger_{\overline{1}'} = U_{\overline{2}'}(U^\dagger_{\overline{2}}U_{\overline{1}})U^\dagger_{\overline{1}'} e^{i\Theta_v} =U^\dagger_{\overline{2}}U_{\overline{2}'}U^\dagger_{\overline{1}'}U_{\overline{1}} e^{i\Theta_v}~.
\end{align}
Hence we have
\begin{align}
U^\dagger_{\overline{2}}U_{\overline{2}'}U^\dagger_{\overline{1}'}U_{\overline{1}} e^{i\Theta_v} = U^\dagger_{\overline{2}}U_{\overline{3}} U^\dagger_{\overline{1}'}U_{\overline{2}'}U_{\overline{3}}^\dagger U_{\overline{1}}~,
\end{align}
which leads to
\begin{align}
    e^{i\Theta_v} =U_{\overline{1}'}U^\dagger_{\overline{2}'}U_{\overline{3}} U^\dagger_{\overline{1}'}U_{\overline{2}'}U_{\overline{3}}^\dagger = e^{i\Theta_{v'}}~.
\end{align}
Therefore, $\Theta_v=\Theta_{v'}$ for an adjacent pair of vertices $v,v'$. This implies that $\Theta_v$ is independent of a vertex $v$,
\begin{align}
    \Theta_v = \Theta~.
\end{align}
This invariant is known to forbid a symmetric SRE state~\cite{Bravyi2006bounds, aharonov2018quantumcircuitdepthlower}, therefore gives another definition of 't Hooft anomaly of $\Z_N$ 1-form symmetry. 

\paragraph{Relation to Else-Nayak type index}
The 't Hooft anomaly of $\Z_N$ 1-form symmetry is characterized by $H^4(B^2\Z_N, U(1)) = \Z_{N\times\gcd(2,N)}$. Each cohomology class is represented by a 4-cocycle 
\begin{align}
    \frac{2\pi i p}{N\times \text{gcd}(2,N)}\mathcal{P}(B) \quad \text{mod $2\pi$}~,
\end{align}
where $p\in \Z_{N\times\gcd(2,N)}$, and $B$ is the 2-form $\Z_N$ background, and $\mathcal{P}(B)$ is Pontryagin square $\mathcal{P}(B):= B\cup B-B\cup_1 dB$ that defines an element of $Z^4(B^2\Z_N, \Z_{N\times \gcd(2,N)})$. The Else-Nayak index $\omega$ corresponds to an element $p\in \Z_{N\times\gcd(2,N)}$. 

Meanwhile, the invariant $\Theta$ is thought to correspond to the spin of the topological line operator for $\Z_N$ 1-form symmetry in continuum QFT, which is given by $\Theta=p/(N\times \text{gcd}(2,N))$. 
Therefore, we conjecture that $\omega$ is in the same class as $2\pi\Theta\mathcal{P}(B)$ in cohomology $H^4(B^2\Z_N, U(1))$.
In Ref.~\cite{kobayashi2025generalizedstatistics, xue2025statistics}, a conjectured correspondence was proposed between such statistical invariants (generalized statistics) and the cohomology group $H^{d+2}(B^{p+1}G, U(1))$ in arbitrary spacetime dimensions. It would be interesting to establish this correspondence generally at the microscopic level, by employing an Else–Nayak type index.

\section{Conclusions}

In this work, we have developed a general framework for characterizing ’t Hooft anomalies of higher-form symmetries in lattice models with tensor product Hilbert spaces. Building on the lattice-based approach of Else and Nayak for 0-form symmetries, we extended the construction to higher-form symmetries by formulating an index valued in group cohomology. In particular, for (2+1) dimensions we defined an index in $H^4(B^2G, U(1))$ associated with Gauss law operators generating a 1-form $G$  symmetry, and we further generalized the construction to arbitrary $(d+1)$ spacetime dimension, where the anomaly is captured by an element of $H^{d+2}(B^{p+1}G, U(1))$. This provides a unified operator-based characterization of higher-form symmetry anomalies in lattice systems and establishes a direct correspondence to their cohomological classification in continuum QFT.
We conclude this paper by listing several possible future directions:
\begin{itemize}
    \item It is recently recognized that 0-form symmetry in a lattice model can have a ``lattice anomaly'' \cite{shirley2025QCA, tu2025anomaliesglobalsymmetrieslattice, kapustin2025anomaly2d}, which is a version of anomalies intrinsic to lattice systems with no counterpart in continuum QFT. For instance, for $G$ 0-form symmetry in (2+1)D, the lattice anomaly is characterized by an index $H^2(BG, \mathbb{Q}_+)$, with $\mathbb{Q}_+$ a GNVW index that characterizes the equivalence class of 1d quantum cellular automata (locality preserving unitaries) \cite{Gross2012}. This index is characterized through the QCA index of a 1d operator $\Omega_{\partial R}:= U_R(g)U_R(h) U_R(gh)^{-1}$ obtained by projective action of symmetries on a disk region $R$. This $H^2$ index becomes an obstruction to onsite realization of symmetry operators, similar to the standard $H^4$ anomaly in (2+1)D.
    Such lattice anomalies are absent for 1-form symmetries in (2+1)D, since the 1d operator $\Omega$ introduced in \eqref{eq:defineOmega} carries the trivial GNVW index. Meanwhile, it is expected that 1-form symmetries in (3+1)D lattice models would exhibit nontrivial lattice anomalies, since the projective action of 1-form symmetry operators on a disk gives a 1d operator which can carry nontrivial GNVW index.~\footnote{Such lattice anomalies are found in (3+1)D $\Z_2$ gauge theory with an emergent fermion, which has $\Z_4$ 1-form symmetry \cite{Barkeshli2023codim2, shirley2025QCA}. } It would be interesting to explore such lattice anomalies for higher-form symmetries. 
    \item For 0-form symmetry in (1+1)D, is known that being free of anomalies is equivalent to being onsite-able \cite{seifnashri2025disentangling}. Meanwhile, higher-form symmetry can be anomalous while being onsite as discussed in Sec.~\ref{subsec:onsite}. It would be interesting to understand the criteria for onsiteability of generic higher-form symmetries~\footnote{Recently, some of the present authors have identified an index that characterizes the obstruction to onsiteability of 1-form symmetries in (3+1)D, taking values in $H^3(B^2G,\mathbb{Q}_+)$~\cite{feng2025onsiteabilityhigherformsymmetries}.}.
    \item It would be interesting to generalize the Else–Nayak type approach to higher-group symmetries \cite{Benini:2018reh}, which naturally appear in lattice models such as the (3+1)D $\mathbb{Z}_2$ toric code \cite{Barkeshli2023codim2, Barkeshli:2022edm, Barkeshli:2023bta}. A promising direction is to characterize higher-group structures via dimensional reduction of symmetry operators. In particular, applying the Else–Nayak procedure to reduce the dimension of a $p$-form symmetry operator may yield a lower-dimensional $q$-form operator ($p<q$) \cite{hsin2025generalizedhallconductivitieslocal}, reflecting the nontrivial mixing of global symmetries of different dimensionalities. In this picture, the anomaly of a $p$-form symmetry could be probed through the anomaly index of the associated $q$-form symmetry obtained by successive reductions. Establishing such a framework would provide a systematic way to capture higher-group structures and their anomalies in lattice models within the Else–Nayak approach.
\end{itemize}

\section*{Acknowledgments}
We thank Maissam Barkeshli, Xie Chen, Meng Cheng, Po-Shen Hsin, Anton Kapustin, and Sahand Seifnashri for stimulating discussions. We thank Anton Kapustin for comments on the draft. 
Y.-A.C. is supported by the National Natural Science Foundation of China (Grant No.~12474491), and the Fundamental Research Funds for the Central Universities, Peking University.
R.K. is supported by the U.S. Department of Energy, Office of Science, Office of High Energy Physics under Award Number DE-SC0009988 and by the Sivian Fund at the Institute for Advanced Study. R.K. thanks the Kavli Institute for Theoretical Physics for hosting the program “Generalized Symmetries in Quantum Field Theory: High Energy Physics, Condensed Matter, and Quantum Gravity” in 2025, during which part of this work was completed. This research was supported in part by grant no. NSF PHY-2309135 to the Kavli Institute for Theoretical Physics (KITP).
S.R. is supported by a Simons Investigator
Grant from the Simons Foundation (Award No.\ 566116).

\appendix

\section{Review: Else-Nayak index for 0-form symmetry in (1+1)D}
\label{app:elsenayak}
Here we review the anomaly index $[\omega]\in H^3(BG,U(1))$ of $G$ symmetry in a (1+1)D lattice model, following Ref.~\cite{ElseNayak2014}.
One begins with a global symmetry operator 
$U(g)$ satisfying the group algebra $U(g)U(h)=U(gh)$ with $g,h\in G$, implemented as a finite-depth circuit in a (1+1)D lattice model. 
For an interval $I$, one defines a restricted operator $U_I(g)$, supported only on $I$. 
Since $U(g)$ is finite-depth, the difference $U_I(g)U_I(h)$ and $U_I(gh)$ is supported at the boundary of $I$. Therefore
\begin{equation}
    U_I(g)\, U_I(h) = \Gamma_{\partial I}(g,h)\, U_I(gh)~,
\end{equation}
where $\Gamma_{\partial I}(g,h)$ is a unitary supported near the endpoints of $I$.

To examine associativity, consider three elements $g,h,k \in G$.  
On one hand,
\begin{equation}
    U_I(g)\, U_I(h)\, U_I(k) \;=\; \Gamma_{\partial I}(g,h)\, \Gamma_{\partial I}(gh,k)\, U_I(ghk)~.
\end{equation}
On the other hand,
\begin{align}
    U_I(g)\, U_I(h)\, U_I(k) 
    &= U_I(g)\, \Gamma_{\partial I}(h,k)\, U_I(hk) \nonumber \\
    &= \big( U_I(g)\, \Gamma_{\partial I}(h,k)\, U_I(g)^{-1} \big)\, U_I(g)\, U_I(hk) \nonumber \\
    &= (^g\Gamma_{\partial I}(h,k))\, \Gamma_{\partial I}(g,hk)\, U_I(ghk)~,
\end{align}
where $^g\Gamma_{\partial I}(h,k)$ denotes the conjugation action of $U_I(g)$ 
on $\Gamma_{\partial I}(h,k)$.
Comparing the two decompositions, one obtains
\begin{equation}
    \Gamma_{\partial I}(g,h)\Gamma_{\partial I}(gh,k) 
    \;=\;(^g \Gamma_{\partial I}(h,k))\Gamma_{\partial I}(g,hk)~.
\end{equation}
Let us denote the two endpoints of an interval $I$ by $l,r$. The operator $\Gamma_{\partial I}$ is given in the form of $\Gamma_{\partial I} = \Gamma_l\Gamma_r$. Then, the associativity of one of the ends $\Gamma_l$ becomes 
\begin{equation}
    \Gamma_{l}(g,h)\, \Gamma_{l}(gh,k) 
    \;=\;  \omega(g,h,k)\, (^g \Gamma_{l}(h,k))\,\Gamma_{l}(g,hk)~,
\end{equation}
where the mismatch $\omega(g,h,k)$ is a $U(1)$ phase.
These phases satisfy the 3-cocycle condition
\begin{equation}
    \omega(g,h,k) \omega(g,hk,l)\omega(h,k,l)\left(\omega( gh,k,l)\omega(g,h,kl)\right)^{-1} = 1~,
\end{equation}
so that $\omega \in Z^3(BG,U(1))$.  
Different choices of $\Gamma_l$ by a phase $\chi(g,h)$ shifts $\omega$ by a coboundary, and 
the anomaly is uniquely characterized by the cohomology class
\begin{equation}
    [\omega]\in  H^3(BG,U(1)).
\end{equation}
This class, which we call the Else-Nayak index, provides a microscopic definition of the 
’t~Hooft anomaly for 0-form $G$ symmetries in (1+1)D lattice models.

\section{Eilenberg-MacLane spaces}
\label{app:EMspace}

In the main text, we claimed that the function $\omega$ generally defined a representative of $H^{d+2}(B^{p+1}G,U(1))$ for $p$-form symmetry in $(d+1)$ spacetime dimensions. In this appendix, we prove this claim by giving a simplicial construction of Eilenberg-MacLane spaces following Chapter V of Ref.~\cite{SimplicialMay1967}. This approach is also mentioned in Ref.~\cite{Kan1958} and Appendix L of Ref.~\cite{Lan2019Fermiondecoration}. 

\subsection{Cohomology of groups}
We begin by reviewing the cohomology of groups. Let us consider a discrete Abelian group $G$. 
For a given $G$ and a $G$-module $M$, define $C^n(G,M)$ to be the $G$-module of all $M$-valued functions of $n$ group elements. In other words, $C^n(G,M)=\{\omega_n\}$, where $\omega_n:G^n\to M$. In the following, we focus on the case where $M$ is the $U(1)$ group and the $G$ action on $M$ is trivial. Also, we use addition to denote the operation in an abelian group, such as $U(1)$ or $G$. 

The abelian groups $C^n(G,U(1))$ can be arranged into a chain complex by introducing a differential operator $\delta_n:C^n(G,U(1))\to C^{n+1}(G,U(1))$ for each $n$. These operators are defined as
\begin{align}
     \nonumber \delta_n\omega_n(g_1,g_2,\cdots,g_{n+1})&=\omega_n(g_2,\cdots,g_{n+1})+(-1)^{n+1}\omega_n(g_1,\cdots,g_n)\\
    &+\sum_{i=1}^{n} (-1)^i\omega_n(g_1,\cdots,g_{i-1},g_i+g_{i+1},\cdots,g_{n+1})~.\label{def of group cohomology diff}
\end{align}
Let $B^n(G,U(1))$ be the image of $\delta_{n-1}$ and let $Z^n(G,U(1))$ be the kernel of $\delta_n$. The group cohomology space is defined as the quotient group
\begin{equation}
    H^n(G,U(1))=Z^n(G,U(1))/B^n(G,U(1))~.
\end{equation}

While group cohomology is a pure algebraic construction, it has another interpretation as the ordinary cohomology of the classifying space $BG$ of group $G$. The space $BG$ has the following simplicial construction. Its simplices are closed configurations of $1$-forms, that is
\begin{equation}
    (BG)_n=Z^1(\Delta^n,G)~,
\end{equation}
where $\Delta^n$ is an $n$-simplex, and $(BG)_n$ is the set of all $n$-simplices of $BG$. The $i$th face of $\alpha\in Z^1(\Delta^n,G)$ is given by its restriction to the $i$th face $\Delta^{n-1}_{(i)}\subset \Delta^n$. In other words, $BG$ is made up of simplices labeled by $1$-cocycles. 

Note that a cocycle $\alpha\in Z^1(\Delta^n,G)$ is completely determined by $\alpha[01],\alpha[12],\cdots,\alpha[n-1\ n]$. Given a cochain $\omega_n^{BG}\in C^n(BG,U(1))$, we construct a cochain $\omega_n^{G}$ of group cohomology by letting
\begin{equation}
    \omega^G_n(g_1,g_2,\cdots,g_n)=\omega^{BG}_n(\alpha) \label{corresponding BG to G}
\end{equation}
such that $\alpha[i-1\ i]=g_i$. We obtain from equation \eqref{corresponding BG to G} a one-to-one correspondence between $C^n(BG,U(1))$ and $C^n(G,U(1))$. Furthermore, we find that the differential operators in \eqref{def of group cohomology diff} correspond to the coboundary operators of $C^*(BG,U(1))$. This proves that the cohomology of group $G$ can be seen as the simplicial cohomology of $BG$. 

\subsection{Simplicial cohomology of Eilenberg-MacLane spaces}
Generalizing the simplicial construction of $BG$, we obtain the following construction of Eilenberg-MacLane spaces $B^{p}G=K(G,p)$. Its simplices are closed configurations of $p$-forms, that is
\begin{equation}
    (B^pG)_n=Z^p(\Delta^n,G)~,
\end{equation}
and the $i$th face of $\alpha\in Z^p(\Delta^n,G)$ is given by its restriction to the $i$th face $\Delta^{n-1}_{(i)}\subset \Delta^n$. In other words, $B^pG$ is made up of simplices labeled by $p$-cocycles. 

From this construction, we obtain that the elements in $C^n(B^pG,U(1))$ are functionals that map each $p$-cocycle labeled $\Delta^n$ to a phase in $U(1)$. In particular, they are of the form $\omega_n(\{g\}_n)$, where $\{g\}_n$ consists of all independent labels on an $n$-simplex. The differential operator has the following form
\begin{equation}
    \delta_n\omega_n(\{g\}_{n+1})=\sum_i (-1)^i \omega_n(\{g\}_n^{(i)})~,
\end{equation}
where $\{g\}_n^{(i)}$ is the restriction of $\{g\}_{n+1}$ to the $i$th face. This operator coincides with the coboundary operator mentioned in Section \ref{Section:Higher form symmetry} if we just consider constant labels. 

Let $B^n(B^pG,U(1))$ be the image of $\delta_{n-1}$ and let $Z^n(B^pG,U(1))$ be the kernel of $\delta_n$. The simplicial cohomology space of $B^pG$ is defined as the quotient group
\begin{equation}
    H^n(B^pG,U(1))=Z^n(B^pG,U(1))/B^n(B^pG,U(1))~.
\end{equation}
It is the generalization of group cohomology to higher-form symmetry. 

\bibliographystyle{utphys}
\bibliography{bibliography}

\end{document}